\documentclass[aps,prx,singlecolumn,groupedaddress,showpacs,superscriptaddress,amssymb,amsmath]{revtex4-2}
\usepackage{graphicx}
\usepackage{cancel}
\usepackage{tabularx}
\usepackage{color}
\usepackage{amsmath}
\usepackage{amsfonts}
\usepackage{amssymb}
\usepackage{comment}
\usepackage{verbatim}
\newcommand{\be}{\begin{equation}}
\newcommand{\ee}{\end{equation}}
\newcommand{\bea}{\begin{eqnarray}}
\newcommand{\eea}{\end{eqnarray}}
\usepackage{dcolumn}
\usepackage{hyperref}
\usepackage{bm}
\usepackage{epsf}
\usepackage{subfig}
\usepackage{epstopdf}
\usepackage{amsfonts}

\newcommand{\unit}{1\!\!1}

\usepackage{amsthm}
\theoremstyle{definition}
\newtheorem{definition}{Definition}[section]

\begin{document}


\title{Equilibrium and nonequilibrium steady states with the repeated interaction protocol: Relaxation dynamics and energetic cost}

\author{Alessandro Prositto}
\affiliation{Department of Physics, University of Toronto, 60 Saint George St., Toronto, Ontario, Canada M5S 1A7}

\author{Madeline Forbes}
\affiliation{Department of Physics, University of Toronto, 60 Saint George St., Toronto, Ontario, Canada M5S 1A7}
\affiliation{Department of Physics and Astronomy, 
University of British Columbia, Vancouver, BC, Canada V6T 1Z1}

\author{Dvira Segal}
\email{dvira.segal@utoronto.ca}
\affiliation{Department of Physics, University of Toronto, 60 Saint George St., Toronto, Ontario, Canada M5S 1A7}
\affiliation{Department of Chemistry and Centre for Quantum Information and Quantum Control,
University of Toronto, 80 Saint George St., Toronto, Ontario, Canada M5S 3H6}

\date{\today}

\begin{abstract}
We study the dynamics of a qubit system interacting with thermalized bath-ancilla spins via a repeated interaction scheme. Considering generic initial conditions for the system and employing a Heisenberg-type interaction between the system and the ancillas, 
we analytically prove the following:
(i) The population and coherences of the system qubit evolve independently toward a nonequilibrium steady-state solution, which is diagonal in the qubit's energy eigenbasis. The population relaxes to this state geometrically, whereas the coherences decay through a more compound behavior.
(ii) In the long time limit, the system approaches a steady state that generally differs from the thermal state of the ancilla. 
We derive this steady-state solution and show its dependence on the interaction parameters and collision frequency.
(iii) We bound the number of interaction steps required to achieve the steady state within a specified error tolerance, 
and we evaluate the energetic cost associated with the process.
Our key finding is that deterministic system-ancilla interactions do not typically result in the system thermalizing to the thermal state of the ancilla. Instead, they generate a distinct nonequilibrium steady state, which we explicitly derive. 
However, we also identify an operational regime that leads to thermalization with a few long and possibly randomized collisions. 
\end{abstract}
\maketitle

\section{Introduction}
\label{sec:intro}

\begin{figure}
  \centering
  \includegraphics[width=0.5\linewidth]{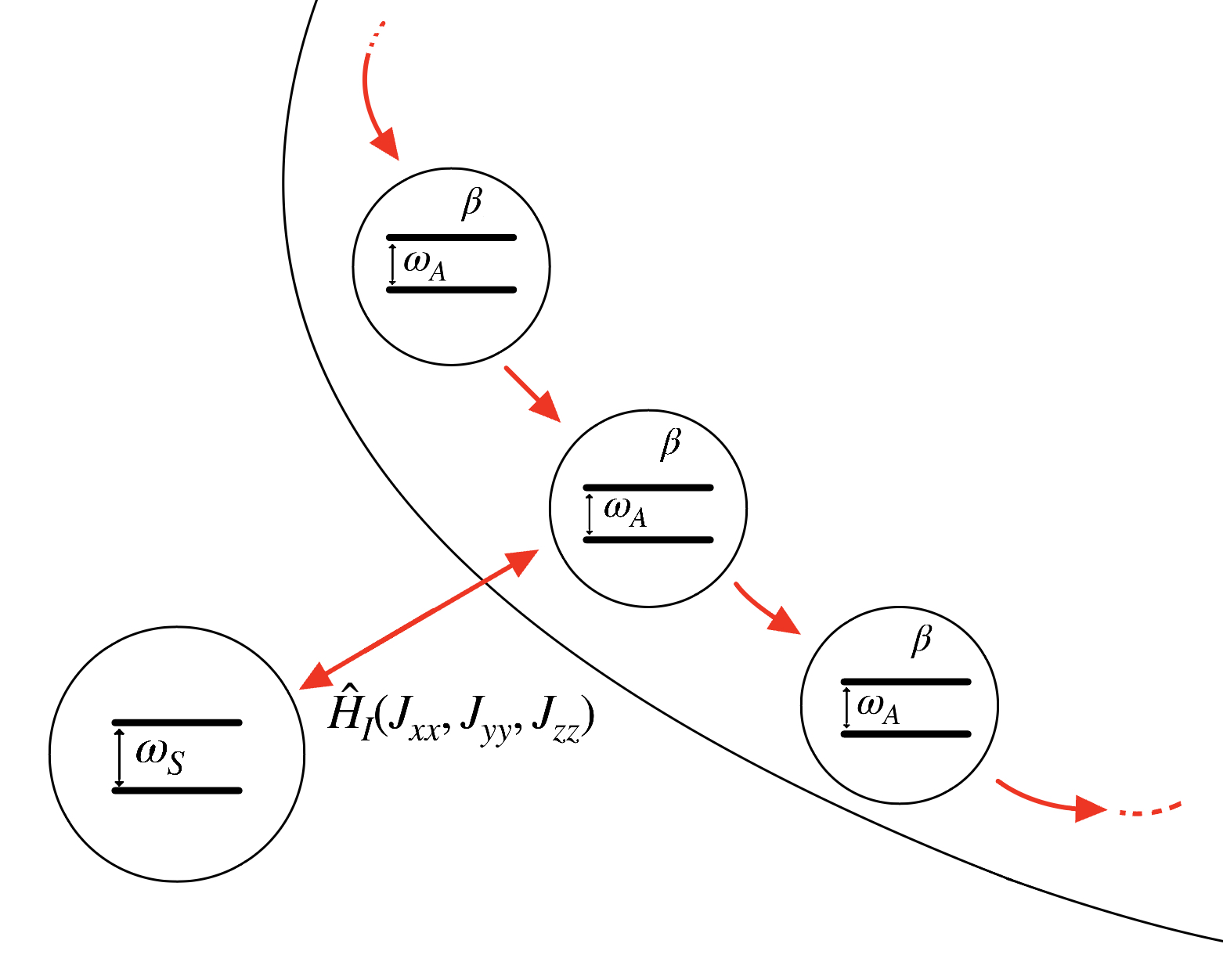}  
   \caption{Scheme of the repeated interaction model under investigation. A system (here a qubit for simplicity) interacts with a collection of independent ancillas for a time interval $\tau$ via the interaction operator $\hat H_I$. After each such collision, the ancilla leaves and a new ``fresh" ancilla instantly arrives to interact with the system.
   }
   \label{fig:figS}
\end{figure}

Repeated Interaction (RI) models have a long history in the theory of open quantum systems \cite{BookOQS}, serving as a framework to model and simulate the interaction of quantum systems with thermal environments, at and beyond the regime of validity of the Lindblad equation \cite{RevRI,brief,Campbell21}.
In essence, the RI approach describes a quantum system that interacts sequentially and repeatedly with many degrees of freedom of a bath. In the basic RI scheme, the bath consists of many identical, noninteracting and uncorrelated spins, commonly referred to as ``ancillas". Each ancilla interacts with the system only once, for a fixed time interval, and it is then discarded, never to interact again; see Fig. \ref{fig:figS}. Alternatively, the RI model can be interpreted as involving a single ancilla, which is refreshed (initialized) after each collision with the system, effectively resetting the ancilla to its initial state before every subsequent collision.

The basic RI model \cite{Gisin2002,Buzek02, Buzek05, Merkli14,Strasberg17} 
has been extended in various directions to address a range of fundamental questions in quantum dynamics and thermodynamics. These extensions include strong coupling thermodynamics \cite{Strasberg19}, finite collision time effects \cite{Scrani19,Campbell21E}, semiclassical effects \cite{Pechen20}, and various types of system-ancilla interactions and ancilla's states, for example, including coherences \cite{Landi19c,Campbell20} and correlations \cite{Landi2021,Filippov22C}.
Other studies focused on the nature of the dynamics captured by RI schemes \cite{Grimmer,Lorenzo_composite,Guarnieri_composite,Zambrini,Layden2016,KosloffE,Landi24,Scarola,Plenio},  
the thermodynamic cost of operating an RI scheme \cite{Barra,Strasberg17,Work22,Parrondo22,YWang22}, its entropy production \cite{Haack24}, and generalizations of the model to real-space scenarios \cite{Parrondo22,BarraQ}. 

The flexibility and ease of implementing the RI scheme, along with its relevance to physical systems such as quantum optics setups, have facilitated its application in a wide range of contexts. Recent studies have employed the RI scheme to achieve thermalization in quantum systems \cite{thermalization,Tabanera-Bravo,Parrondo22,BarraPRX21}, generate a nonequilibrium steady state in spin chains \cite{XXchain}, and produce a nonequilibrium steady state with quantum coherences \cite{Qcoh,Chiara24}. Quantum technology applications that involve the RI scheme include proposals for quantum thermometry \cite{thermometry1,thermometry2}, designs for quantum batteries \cite{Battery21,Barra22}  and studies of quantum thermal machines \cite{machine, refri,Goold22P,Goold23Heat}. Furthermore, the RI framework is attractive for performing quantum simulations of open quantum systems, and it has been proposed as a practical tool for implementations on quantum computing hardware, both in near-term devices \cite{Poletti23,Donadi,Vedral} and in fault-tolerant quantum computers \cite{MattRI}.

In particular, the RI framework has been employed to investigate two significant processes in quantum information theory: quantum homogenization \cite{Gisin2002} and quantum thermalization \cite{Buzek02}. Quantum homogenization describes the process by which a system, initially in an arbitrary state, interacts with a reservoir and evolves toward a steady state, becoming arbitrarily close --- but not identical 
\cite{Zurek1982, Zurek2009} ---  to the state of the reservoir \cite{Gisin2002, Buzek05}.
Quantum thermalization, while closely related to quantum homogenization, differs in that the reservoir's degrees of freedom (ancillas) are in a Gibbs thermal state. Furthermore, it is assumed that the reservoir possesses a sufficiently large number of degrees of freedom, ensuring that any changes to its state during interactions remain negligible \cite{Gisin2002, Buzek02}.
Quantum thermalization via an RI scheme finds applications in a wide range of fields, including quantum many-body physics \cite{Archak,Archak23}, quantum computing \cite{Beever2024}, and quantum cryptography \cite{Gisin2002, Buzek02}. In quantum many-body systems, thermalization plays a critical role in understanding how closed systems evolve toward equilibrium, bridging the gap between microscopic quantum dynamics and macroscopic thermodynamics. Recent advances have extended the study of thermalization and homogenization in collisonal models to non-Markovian scenarios, where system-reservoir interactions exhibit memory effects \cite{Giovannetti,Ciccarello_Palma, Lorenzo_MemoryKernel,Kretschmer, Landi2021, Ghosh2024}. 


In the seminal works \cite{Gisin2002, Buzek02}, it was analytically demonstrated that a partial swap is the only unitary operation capable of achieving both quantum homogenization and thermalization, regardless of the initial state of the system or the reservoir's ancillas. This process ensures a monotonic convergence of the system's state toward that of the ancillas. A partial swap can be shown to correspond to the Heisenberg exchange interaction Hamiltonian with {\it identical} coupling constants for the $x$ and $y$ components \cite{Buzek02, Vedral, Fan2005}. The case of {\it asymmetry} in the Heisenberg interaction Hamiltonian has been explored numerically in Ref. \citenum{Campbell20}. Additionally, Ref. \citenum{Gisin2002} establishes a lower bound on the number of iterations required to achieve homogenization or thermalization within a specified accuracy. Furthermore, Ref. \citenum{Beever2024} examines the capacity of a reservoir to carry out multiple homogenization or thermalization processes.

With this significant activity on the capacity, extensions, and applications of the RI scheme, it is surprising that basic questions about the steady state achieved within the RI framework remain unanswered. Here, we focus on the most basic RI model: a qubit system interacting with a bath of uncorrelated and noninteracting spin ancillas. In this context, we consider the rich scenario where (i) the system-ancilla interactions can be of an arbitrary strength, (ii) the interaction duration is not constrained to be short, relaxing the conventional assumption of brief interactions, and (iii) the interactions extend beyond the ``energy conserving" model (partial swap), incorporating the general form of the Heisenberg interaction.

Assuming the bath ancillas are prepared in a thermal state, the central focus of our study is {\it whether---and to what state---the system relaxes}. This question has previously been explored in Ref. \citenum{RevRI}, particularly through numerical simulations as in Ref. \citenum{Campbell20}. The primary conclusion from those studies is that the system {\it thermalizes} to the state of the bath ancillas {\it only} if the interactions are of an ``energy conserving" nature. 
Beyond this, numerical studies have uncovered a range of outcomes: depending on the interaction Hamiltonian, the system may converge to either a diagonal or a nondiagonal steady state. Remarkably, in some interaction models, the steady state depends on the initial conditions of the system, even when the bath ancillas are prepared in a canonical thermal state \cite{Campbell20}. 

Our study addresses some of these observations with exact analytical work: Assuming generic initial conditions, we derive the relaxation dynamics and the eventual nonequilibrium steady state of the system, which is generally nonthermal. We identify, however, a protocol under which an RI scheme with interactions beyond a partial swap leads to an actual qubit thermalization, and in only a few RI steps.


It is important to remember that each iteration of the RI protocol incurs an energetic cost associated with the work required to switch the interaction between the system and the ancilla on and off \cite{Barra}. With the objective to create a certain steady state, it is necessary to quantify the number of RI iterations needed to bring the system, initialized to some state, to the target steady state within a certain distance metric, such as the trace distance or fidelity. 
Once the required number of iterations necessary to approach the steady state is determined, the total work investment necessary to create the steady state can be calculated. This consideration is particularly important for applications in quantum computing and quantum technologies, where energy efficiency and resource optimization play a critical role \cite{Auffeves}. 


For two interaction models termed the $J_{xx}-J_{yy}$ and $J_{xx}-J_{yy}-J_{zz}$ models,
we solve the problem of creating a steady state exactly and analytically, starting from a generic initial state for the system. We further complement analytical work with simulations as necessary.
Our main contributions are as follows.

(i) {\bf Relaxation dynamics:} For both models, we show that the population and coherences evolve {\it independently} to their steady state, which is diagonal in the energy basis of the qubit. While population relaxes to the steady state geometrically, coherences decay (to zero) through a more compound behavior. 
%

(ii) {\bf Nonequilibrium steady state:} We derive explicit expressions for steady-state population of the system, a global fixed-point of the evolution map, which is independent of initial conditions. We show that in general the steady-state population depends not only on the temperature of the ancilla and the energy structure of the system, as in standard thermalization, but also on the form of the system-ancilla interaction Hamiltonian and its parameters, and even on the interaction duration. 
%

(iii) {\bf Thermalization in a few steps:} Based on our analytical results, we identify and exemplify an RI protocol that achieves thermalization of the qubit with the ancillas. This RI protocol requires long collisions, yet small (possibly random) interaction strengths.

(iv) {\bf Protocol runtime:} For diagonal initial states of the system, we derive a closed-form expression for the number of RI collisions required to get $\epsilon$ close (via a trace distance metric) to the target state. This number scales logarithmically with the desired accuracy and approximately inversely with the square of the interaction energy.

(v) {\bf Energetic cost:} We derive and simulate the work required in the RI process to generate a steady state, considering various initial conditions and for different interaction models. Our results reveal that, besides the special point $J_{xx}=J_{yy}$ where thermalization occurs as a heat transfer process without consuming additional energy,
work is required to reach the steady state when $J_{xx}\neq J_{yy}$.
We also prove that erasing coherences in our models does not require energy exchange between the system and the ancillas.


The paper is organized as follows. In Sec. \ref{sec:XY}, we analyze the nonequilibrium steady state of the $J_{xx}-J_{yy}$ model through the RI scheme. Sec. \ref{sec:XYZ} describes the more general relaxation dynamics in the $J_{xx}-J_{yy}-J_{zz}$ interaction model. 
In Sec. \ref{sec:random}, we discuss a protocol that achieves thermalization in our models.
We turn to resource cost estimation in Sec. \ref{sec:energetics}, and we summarize with some open questions in Sec. \ref{sec:summary}.

\section{RI dynamics and  steady state: $J_{xx}-J_{yy}$ model}
\label{sec:XY}

\subsection{Model}

In the RI scheme, the system repeatedly interacts with refreshed spins (ancillas) of the environment, each prepared in a thermal state $\rho_A$. For a schematic representation of the model, see Fig. \ref{fig:figS}.
The free Hamiltonians of the system $S$ and the ancilla $A$ are given by
\bea
    \hat{H}_{S} = -\frac{\omega_{S}}{2}\hat{\sigma}_{z}^{S}, \,\,\,\,\,\,
    \hat{H}_{A} = -\frac{\omega_{A}}{2}\hat{\sigma}_{z}^{A},
\label{eq:HSHA}
\eea
where $\hat{\sigma}_{z}^{\bullet}$ is the $z$-component of Pauli spin operators and $\omega_{S}$ and $\omega_{A}$ are the energy splitting of the  system and the ancilla, respectively. In general, we allow these splitting to be distinct. As for the system-ancilla interaction Hamiltonian, $\hat{H}_{I}$, we assume the form
\begin{equation}
    \hat{H}_{I} = J_{xx}\hat{\sigma}_{x}^{S}\otimes \hat{\sigma}_{x}^{A} +J_{yy}\hat{\sigma}_{y}^{S}\otimes \hat{\sigma}_{y}^{A},
    \label{eq:HI_JxxJyy}
\end{equation}
where $\hat{\sigma}_{x}^{\bullet}$ and $\hat{\sigma}_{y}^{\bullet}$ are the $x$ and $y$ components of Pauli spin operators, respectively, and $J_{xx}$ and $J_{yy}$ are the coupling strengths along $x$ and $y$, respectively. In Sec. \ref{sec:XYZ}, we include additional system-ancilla interactions, $J_{zz}\hat\sigma_z^S\otimes\hat\sigma_z^A$, completing the Heisenberg-type interaction model.

During each RI step, the total Hamiltonian is given by
\begin{equation}
   \hat{H}_{\text{tot}} = \hat{H}_{S}\otimes \unit^{A}+\unit^{S}\otimes\hat{H}_{A} + \hat{H}_{I},
\label{eq:Htot}
\end{equation}
where the symbol $\unit^{\bullet}$ indicates the identity operator in the Hilbert space of the system $S$ or the ancilla $A$.

Before a collision with the system, the state of each ancilla, $\rho_{A}$, is assumed to be a Gibbs thermal state prepared at the inverse temperature of the bath, $\beta$, namely,
\begin{equation}
\label{AncillaInitialState}
\rho_{A} =  \begin{pmatrix}
       p_{A} & c_{A} \\
       c_{A}^{*} & 1-p_{A}
\end{pmatrix} = 
\begin{pmatrix}
\frac{1}{1+e^{-\beta\omega_{A}}} & 0 \\
0 & \frac{e^{-\beta\omega_{A}}}{1+e^{-\beta\omega_{A}}}
\end{pmatrix}.
\end{equation}
Here, $p_A$ is the population of the ground state of the ancilla, 
dictated by the canonical distribution,
while $c_A$ are the coherences, which we assume are null.
We next study the dynamics of the qubit system through the RI scheme, starting from a generic initial condition and reaching the steady state.

\subsection{Dynamics}
The initial state of the system $S$, denoted by $\rho_S^{(0)}$, can be chosen at random. In the $n$th iteration of the RI protocol, the state of the system,  $\rho_{S}^{(n)}$, will take the general form
\begin{equation}
   \rho_{S}^{(n)} = \begin{pmatrix}
       p_{S}^{(n)} & c_{S}^{(n)} \\
       (c_{S}^{(n)})^{*} & 1-p_{S}^{(n)}.
\end{pmatrix}
\end{equation}
For later use, we also {\it define} an effective temperature for the system, which is meaningful for diagonal states, as $\beta_S^{(n)} \equiv -\frac{1}{\omega_S} \ln \left(\frac{1-p_S^{(n)}}{p_S^{(n)}}\right)$. 

We now describe the time evolution in the RI scheme. 
At each collision, the collision unitary $\hat{U}(\tau)$ is given by
 $\hat{U}(\tau) = e^{-i\hat{H}_{\text{tot}}\tau}$, where $i$ is the imaginary unit, $\tau$ is the collision time, namely the duration of each collision, and we have set $\hbar =1$.
At each RI step, the dynamics of the system is described by the equation
\bea 
\rho_S^{(n+1)}& = & {\rm Tr_A} \left[ \rho_{\text{tot}}^{(n+1)}\right]    
\nonumber\\
&=&
{\rm Tr_A}
\left[ \hat{U}(\tau)\rho_{\text{tot}}^{(n)} \hat{U}^{\dagger}(\tau)\right] 
 \nonumber\\
& =&
   {\rm Tr_A}
\left[\hat{U}(\tau) \left(\rho_{S}^{(n)}\otimes\rho_{A}\right)\hat{U}^{\dagger}(\tau)\right],
\label{eq:RI}
\eea
where $\rho_{\text{tot}}^{(n)}$ is the state of the total system after $n$ collisions, and $\rho_{\text{tot}}^{(n+1)}$ is the resulting state after one more collision. Since the interaction Hamiltonian $\hat{H}_{I}$ and the collision time $\tau$ are fixed, the collision unitary $\hat{U}(\tau)$ is the same for each collision.
At each RI time step, we tensor-product the system's state with a {\it refreshed} ancilla, assuming no correlations between them or between fresh and used ancillas.
The density matrix of the system after $n$ collisions, $\rho_{S}^{(n)}$, is thus obtained by tracing $\rho_{\text{tot}}^{(n)}$ over the degrees of freedom of the ancilla \cite{PT}. Note that we assume that there is no idle time between collisions. That is, the full time interval is divided into a set of RI collisions, each of duration $\tau$.
In a matrix form, the total Hamiltonian (\ref{eq:Htot}) is written as
\begin{equation}
\hat{H}_{\text{tot}} =
\begin{pmatrix}
 -\frac{\omega _A}{2}-\frac{\omega _S}{2} & 0 & 0 & J_{xx}-J_{yy} \\[0.25cm] 
 0 & \frac{\omega _A}{2}-\frac{\omega _S}{2} & J_{xx}+J_{yy} & 0 \\[0.25cm] 
 0 & J_{xx}+J_{yy} & -\frac{\omega _A}{2}+\frac{\omega _S}{2} & 0 \\[0.25cm] 
 J_{xx}-J_{yy} & 0 & 0 & \frac{\omega _A}{2}+\frac{\omega _S}{2} \\[0.25cm] 
\end{pmatrix},
\end{equation}
with the four basis states corresponding to the spin state in the system and the ancilla, $|\downarrow_S \downarrow_A \rangle$, $|\downarrow_S \uparrow_A \rangle$, 
$|\uparrow_S \downarrow_A  \rangle$, $|\uparrow_S \uparrow_A \rangle$.
This results in the collision unitary matrix,
\begin{widetext}
\begin{equation}
\label{eq:U}
\small
\hspace*{-2.0cm}
\hat{U}(\tau) = 
\begin{pmatrix}
\cos \left(\frac{\phi\tau }{2}\right)+i\frac{\left(\omega _A+\omega _S\right)}{\phi}\sin \left(\frac{\phi\tau }{2}\right) & 0 & 0 & -2 i\frac{ \left(J_{xx}-J_{yy}\right) }{\phi}\sin \left(\frac{\phi\tau }{2}\right) \\
 0 & \cos \left(\frac{\theta  \tau }{2}\right)-i\frac{ \left(\omega _A-\omega _S\right) }{\theta}\sin \left(\frac{\theta  \tau }{2}\right) & -2 i \frac{  \left(J_{xx}+J_{yy}\right)}{\theta}\sin \left(\frac{\theta  \tau }{2}\right) & 0 \\
 0 & -2 i \frac{\left(J_{xx}+J_{yy}\right)}{\theta}\sin \left(\frac{\theta  \tau }{2}\right) & \cos \left(\frac{\theta  \tau }{2}\right)+i\frac{ \left(\omega _A-\omega _S\right) }{\theta}\sin \left(\frac{\theta  \tau }{2}\right) & 0 \\
 -2 i \frac{\left(J_{xx}-J_{yy}\right) }{\phi}\sin \left(\frac{\phi\tau }{2}\right) & 0 & 0 & \cos \left(\frac{\phi\tau }{2}\right)-i\frac{ \left(\omega _A+\omega _S\right) }{\phi}\sin \left(\frac{\phi\tau }{2}\right) \\
\end{pmatrix} 
\end{equation}
\end{widetext}
where we defined the energy parameters $\theta$ and $\phi$ as
\bea
\label{eq:thetaphi}
    \theta &=& \sqrt{4(J_{xx}+J_{yy})^{2} + (\omega_{A}-\omega_{S})^2},
\nonumber\\
    \phi &=& \sqrt{4(J_{xx}-J_{yy})^{2} + (\omega_{A}+\omega_{S})^2}.
\eea
Substituting the expression for the time evolution operator (\ref{eq:U}) into Eq. (\ref{eq:RI}),  we find that the population and coherences are decoupled in their evolution through the RI map,
that is, $p_S^{(n)}$ depends only on $p_S^{(n-1)}$, and not on $c_S^{(n)}$ or $c_S^{(n-1)}$, and similarly for the coherences.
We next separately analyze their evolution through repeated interactions.

\subsection{Population dynamics: Analytical results and simulations}

We simplify the RI step, the result of Eq. (\ref{eq:RI}), finding a seemingly cumbersome result,
\bea
\label{eq:pop}
\notag p_{S}^{(n+1)} &=&
 4\frac{(J_{xx}-J_{yy}){}^2}{\phi ^2}(1-p_{A})\sin ^2\left(\frac{\phi\tau }{2}\right) + 
 4p_{A}\frac{ (J_{xx}+J_{yy} ){}^2}{\theta ^2}\sin ^2 \left(\frac{\theta\tau}{2}\right) \nonumber \\ &+&
  \notag  p_{S}^{(n)} \Biggl[\frac{1}{2} \left(1-p_{A}\right) (1+\cos (\theta  \tau )) - 
  4 \left(1-p_{A}\right)\frac{ \left(J_{xx}-J_{yy}\right){}^2 }{\phi ^2}\sin ^2\left(\frac{\phi\tau }{2}\right) \nonumber \\ &+& \notag  p_A \cos ^2\left(\frac{\phi\tau }{2}\right) -4 p_A  \frac{\left(J_{xx}+J_{yy}\right){}^2 }{\theta ^2}\sin ^2\left(\frac{\theta  \tau }{2}\right) + \left(1-p_A\right)\frac{ \left(\omega _A-\omega _S\right){}^2 }{\theta ^2}\sin ^2\left(\frac{\theta  \tau }{2}\right)  \nonumber\\
  &+& p_A\frac{ \left(\omega _A+\omega _S\right){}^2 }{\phi ^2}\sin ^2\left(\frac{\phi\tau }{2}\right)\Biggr].
\eea
This expression was derived by 
identifying the ancilla's excited state population, 
 $1-p_A=\frac{1}{1+e^{\beta  \omega _A}}$ and its ground state population,
 $p_A=\frac{1}{1+e^{-\beta \omega _A}}$. We also use $\theta$ and $\phi$ as energy parameters given by Eq. (\ref{eq:thetaphi}).
 
Equation (\ref{eq:pop}) describes a single RI step. It is exact, but it does not provide much insight as is. 
Specifically, the challenge is to use theis expression and build the steady-state solution.
We now describe how we find it. We put forward the following ansatz (guess) for the evolution of populations:
\begin{equation}
\label{eq:ansatz}
    p_{S}^{(n+1)}-p_{S}^{(\infty)} = \eta\left(p_{S}^{(n)}-p_{S}^{(\infty)}\right).
\end{equation}
Here, $p_{S}^{(\infty)}$ is the (yet unknown) ground-state population of the system $S$ in the steady-state (long-time) limit. The coefficient $\eta$ is a function of microscopic parameters, and it dictates the rate of convergence to the steady-state solution.
Looking at Eq. (\ref{eq:pop}), we construct $\eta$ by finding the coefficient of $p_S^{(n)}$ from the relation $p_{S}^{(n+1)} = \eta p_{S}^{(n)} +C$, with $C$ as leftover terms that do not depend on $p_S^{(n)}$.
Next, we propose that the constant $C$ is given by $C=p_S^{(\infty)}(1-\eta)$. We extract the asymptotic solution, $p_S^{(\infty)}$, from this relation based on the previously received $\eta$.
This procedure allows us to identify the rate $\eta$ from Eq. (\ref{eq:pop}),  
%
\bea
\label{eq:eta}
 \eta & =& \frac{\left(\omega _A-\omega _S\right){}^2 }{\theta ^2}\sin ^2\left(\frac{\theta  \tau }{2}\right)+\frac{1}{2} (1+\cos (\theta  \tau ))-4\frac{ \left(J_{xx}-J_{yy}\right){}^2 }{\phi ^2}\sin ^2\left(\frac{\phi\tau}{2}\right) 
\nonumber\\
&+&p_{A}\Biggl[-\frac{1}{2} (1+\cos (\theta  \tau ))+\cos ^2\left(\frac{\phi\tau}{2}\right)+4\frac{ \left(J_{xx}-J_{yy}\right){}^2 }{\phi ^2}\sin ^2\left(\frac{\phi\tau }{2}\right)  
\nonumber\\ 
&-&\frac{\left(\omega _A-\omega _S\right){}^2 }{\theta ^2}\sin ^2\left(\frac{\theta  \tau }{2}\right)+\frac{\left(\omega _A+\omega _S\right){}^2 }{\phi ^2}\sin ^2\left(\frac{\phi\tau }{2}\right)-4\frac{ \left(J_{xx}+J_{yy}\right){}^2}{\theta ^2}\sin ^2\left(\frac{\theta  \tau }{2}\right) \Biggr],
\eea
which after simplifications, yields the elegant result
\begin{equation}
    \label{eq:eta2}
    \eta = 1
       - \frac{4(J_{xx}+J_{yy})^2}{\theta^{2}} \sin^{2}\left(\frac{\theta\tau}{2}\right) 
       - \frac{4(J_{xx}-J_{yy})^2}{\phi^{2}}\sin^{2}\left(\frac{\phi\tau}{2}\right).
\end{equation}
Remarkably, this rate of building the long-time solution: (i) does not depend on the temperature of the ancilla,
(ii) varies nonmonotonically with the collision time $\tau$, and (iii) depends on the system-ancilla interaction parameters $J_{xx}$ and $J_{yy}$ as well as the system and ancilla's frequencies, $\omega_{S}$ and $\omega_{A}$, respectively.
(iv) It is also straightforward to prove that $-1<\eta<1$, except for a countable set of points where $\eta$ reaches the value $1$. 
These lower and upper bounds ensure otherwise convergence to a stationary limit. For more details, see Appendix \ref{appendA}.

From the ansatz (\ref{eq:ansatz}) we also construct the steady-state solution of the ground state population, $p_{S}^{(\infty)}$, as
\begin{equation}
\label{eq:psss}
\hspace{-2cm}
p_{S}^{(\infty)} = 
\frac{\theta^{2}(1-\cos(\phi\tau))(J_{xx}-J_{yy})^{2}(1-p_{A}) + \phi^{2}(1-\cos(\theta\tau))(J_{xx}+J_{yy})^{2}p_{A}}
{\phi^{2}(1-\cos(\theta\tau))(J_{xx}+J_{yy})^2+\theta^{2}(1-\cos(\phi\tau))(J_{xx}-J_{yy})^{2}}.
\end{equation}
This expression, which represents one of our main results, explicitly demonstrates that in a general RI scheme with a collision time step $\tau$ and interaction couplings $J_{xx}\neq J_{yy}$, the system generically {\it does not} thermalize to the state of the environmental ancillas. 
Importantly, the dependence of the steady-state population on the collision time $\tau$ becomes negligible in the limit of sufficiently short collision time, as we show next. We now discuss some significant limits of Eq. (\ref{eq:psss}):

(1) When $J_{xx}\rightarrow J_{yy} \equiv J$, the system converges to the ancillary populations, $p_S^{(\infty)}\to p_A$, regardless of the time step $\tau$, the difference in frequencies $\omega_{A}-\omega_S$, and the interaction energy $J$.
%
%
The relaxation rate depends on the energy and timescale of the interaction according to $\eta=1-16(J^2/\theta^2)\sin^2(\theta\tau/2)$.
The $J_{xx}=J_{yy}$ limit in fact corresponds to ``energy conserving" interactions, for which no work is done during an RI step \cite{RevRI,Campbell20}.
The equivalence of population $p_S^{(\infty)}=p_A$ for $J_{xx}=J_{yy}$ translates to the relation $\beta_S^{(\infty)}\omega_S = \beta\omega_A$. Therefore, 
to thermalize the system to the actual temperature of the ancilla, $1/\beta$, one needs to set $\omega_S=\omega_A$ \cite{commentT}.



(2) In the limit of infinitesimal collision time, $\tau\rightarrow0^{+}$, it is possible to verify from Eq. (\ref{eq:eta2}) that $\eta\rightarrow1$.  Thus, as trivially expected, our ansatz yields $p_{S}^{(n+1)}\rightarrow p_{S}^{(n)}$. The reasoning behind this is that if the duration of interactions between the system and the environmental ancilla are ultra-short, there is insufficient time for information to transfer from the ancilla to the system. Consequently, the population of the system remains fixed at its initial condition.

(3) More relevant is the solution for short yet finite $\tau$. We then expand both $\eta$ and $p_{S}^{(\infty)}$ as series of $\tau$ around $\tau=0$. The first nontrivial second-order term yields
\bea
p_{S}^{(\infty)}& =&\frac{ 4 p_A J_{xx} J_{yy}+\left(J_{xx}-J_{yy}\right){}^2}{2(J_{xx}^{2}+J_{yy}^{2})} + O(\tau^4),
\nonumber\\
\eta&=&1-2(J_{xx}^2+J_{yy}^2)\tau^2 + O(\tau^4).
\label{eq:ptauss}
\eea
One can also express the population in Eq. (\ref{eq:ptauss}) in a form that highlights deviations from the state of the ancilla,
\bea
p_{S}^{(\infty)}& =&p_A + (1-2p_A)\frac{ \left(J_{xx} - J_{yy}\right)^2}{2\left(J_{xx}^{2}+J_{yy}^{2}\right)} + O(\tau^4).
\eea
It is now clear that when $J_{xx}=-J_{yy}$, the ground state of the system reaches the population $p_S^{(\infty)}=(1-p_A)$, that is, we observe a {\it complete} population inversion between the states, compared to the state of the ancilla. It can be proved from Eq. (\ref{eq:ptauss})
that population inversion takes place whenever the signs of $J_{xx}$ and $J_{yy}$ differ (assuming a thermal ancilla, $p_A\geq 1/2$), irrespective of the magnitude of the coupling energies; the effect of population inversion can be represented as a negative temperature for the system \cite{Campbell20}. 

(4) Another interesting limit of Eq. (\ref{eq:ptauss}) arises when keeping only one of the interaction terms nonzero, e.g., $J_{xx}\neq0$
but $J_{yy}=0$, and when the collision time $\tau $ is short such that we expand the trigonometric functions in series at $2^{nd}$ order in $\tau$ around 0.  In this case, $p_S^{(\infty)}\to 1/2$, an effective infinite temperature situation.

\begin{figure}
    \centering
\includegraphics[width=0.75\linewidth]{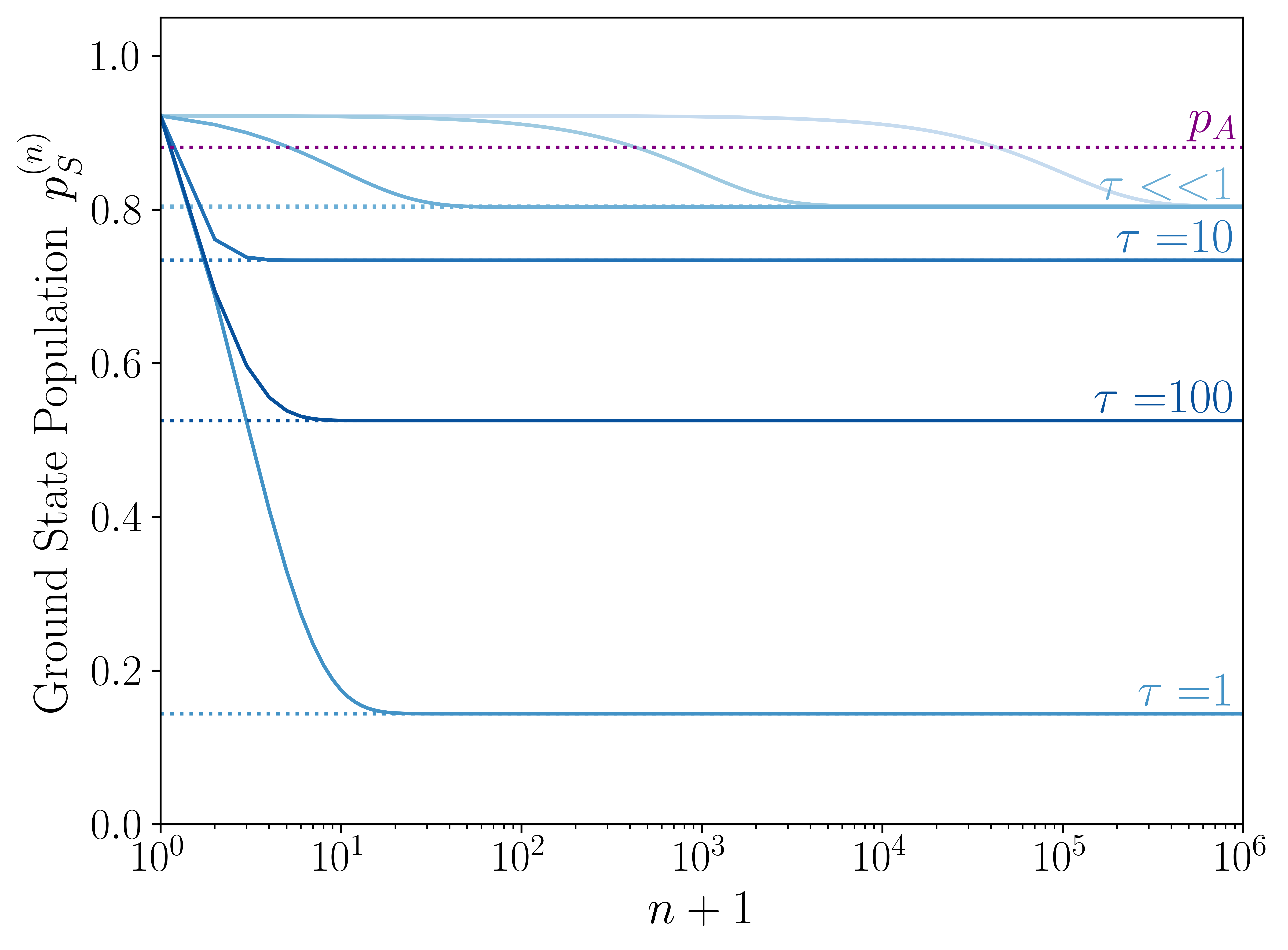} 
\caption{Evolution of the ground state population of the system $S$ for the $J_{xx}-J_{yy}$ interaction model. The dynamics is simulated with $10^{6}$ time steps, taking $\omega_{A} = 2$,  $\omega_{S} = 1$, $J_{xx} = 2$, $J_{yy}=1$, $J_z=0$, $\beta = 1$. 
Simulation are repeated for different collision times,
$\tau\in\{10^{-3},10^{-2},10^{-1},1,10,10^{2}\}$ as marked in the figure, light to dark, with RI simulations (solids) and predictions from the steady state equation (\ref{eq:psss}) (dotted). 
The initial state of the system was taken at random, but we use the same choice for all values of $\tau$. 
We also mark the ground state population of the environmental ancillas, $p_A$, prepared in a thermal Gibbs state  at the inverse temperature $\beta=1$.
For presentation reasons, here the initial state is set as $n=1$. 
}
 \label{figP}
\end{figure}

In sum, we have shown that the population evolves independently of coherences, and that we can write $p_{S}^{(n+1)}$ in the recursive form of Eq. (\ref{eq:ansatz}), where $p_{S}^{(\infty)}$ is given by Eq. (\ref{eq:psss}) and the rate is given by Eq. (\ref{eq:eta2}). Iterating the dynamics $n$ times, it is clear that $p_{S}^{(n)}$ converges to $p_{S}^{(\infty)}$ geometrically as
\begin{equation}
    p_{S}^{(n)}-p_{S}^{(\infty)} = \eta^n\left(p_{S}^{(0)}-p_{S}^{(\infty)}\right).
\label{eq:geom}
\end{equation}
We now illustrate these findings through simulations and compare them with our analytical results. 
Figure \ref{figP} presents the evolution of the ground state population of the reduced system $S$, plotted as a function of the number of repeated interaction steps, $n$. The different lines correspond to different choices of the time step $\tau$, and we compare RI simulations (full) to steady-state predictions (dotted), which are based on Eq. (\ref{eq:psss}). 
%

In perfect agreement with our predictions, we find that the steady state of the system generally depends on $\tau$; population approaches a $\tau$-independent value only once $\theta\tau\ll1$ and $\phi\tau\ll1$. 
Importantly, the dependence of the steady-state on the duration $\tau$ is nonmonotonic, which can be understood from Eq. (\ref{eq:psss}) as resulting from oscillatory coherent dynamics during the interaction time.
Furthermore, since we work with parameters such that  $J_{xx}\neq J_{yy}$, the system's ground state does not approach the ground state population of the ancilla, marked by $p_A$ in Fig. \ref{figP}.

\begin{figure}
    \centering
    \includegraphics[width=0.5\linewidth]{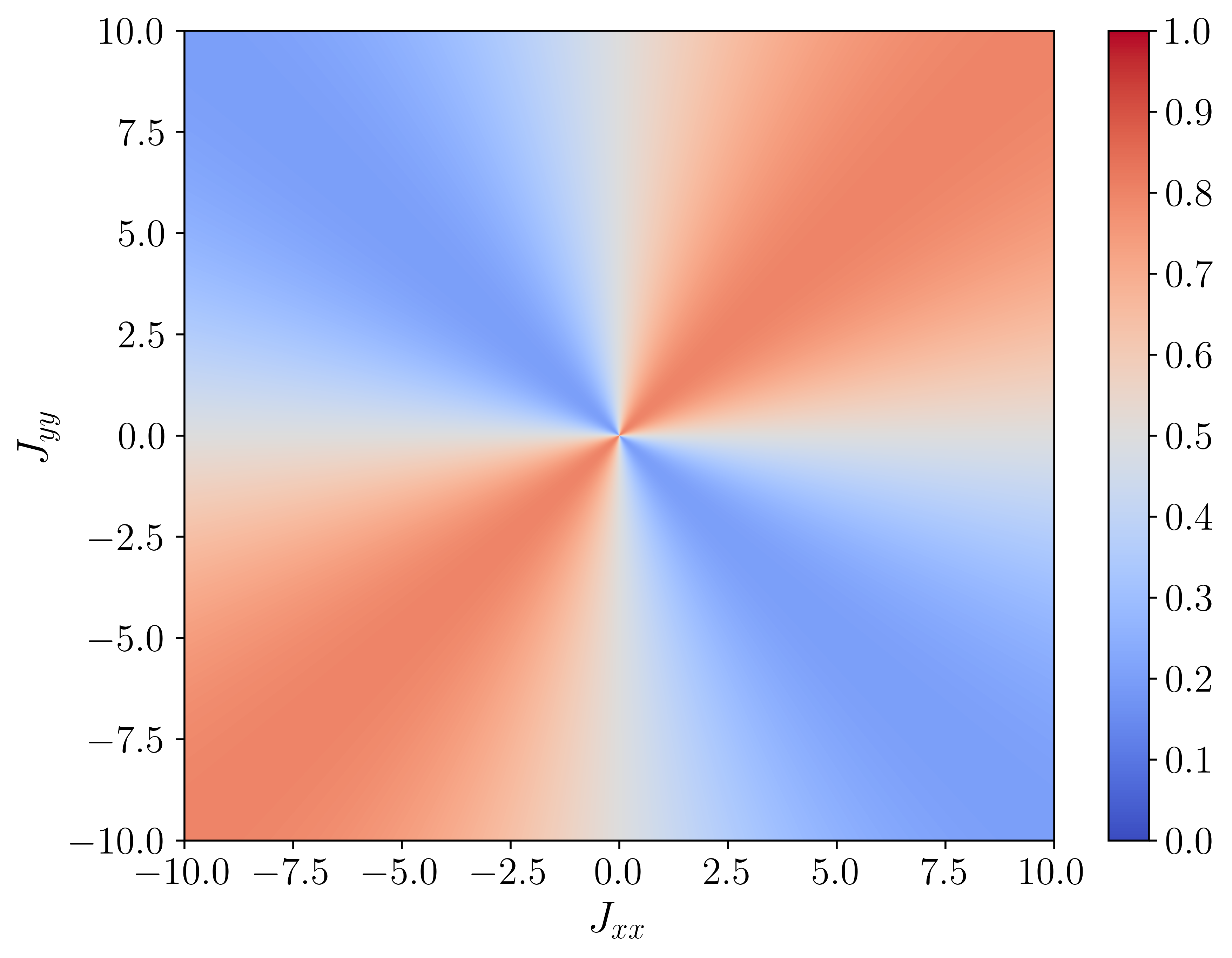}
\caption{
Contour plot of the steady-state (long time) ground state population of the system $S$ predicted under the RI scheme presented here.
Populations are shown as a function of $J_{xx}$ and $J_{yy}$. 
Parameters are $\tau=0.01$,  $\omega_A=\omega_S=1$, $p_A=0.8$.
}
    \label{figM}
\end{figure}

In Figure \ref{figM}, we present a contour plot of the steady-state population of the system, $p_S^{(\infty)}$, presented as a function of both $J_{xx}$ and $J_{yy}$. This calculation, based on Eq. (\ref{eq:psss}, was done using a short collision time, $\tau = 0.01$; in fact, for this value, steady-state results are independent of $\tau$, as we demonstrated in Fig. \ref{figP}.
As predicted by Eq. (\ref{eq:ptauss}), we observe an odd symmetry of the steady state with respect to $J_{xx}\leftrightarrow J_{yy}$. Furthermore, when $J_{xx}=J_{yy}$, $p_S^{(\infty)}=0.8$, which corresponds to the population of the ground state of the ancilla; when $J_{xx}=-J_{yy}$, $p_S^{(\infty)}=1-p_A=0.2$, that is, we observe complete population inversion with respect to the ancilla, according to Eq. (\ref{eq:ptauss}). In fact, whenever the signs of $J_{xx}$ and $J_{yy}$ are opposite, population inversion occurs. 

\subsection{Coherence dynamics: Analytical results and simulations}

\begin{figure}[htbp]
  \centering
  \includegraphics[width=0.75\linewidth]{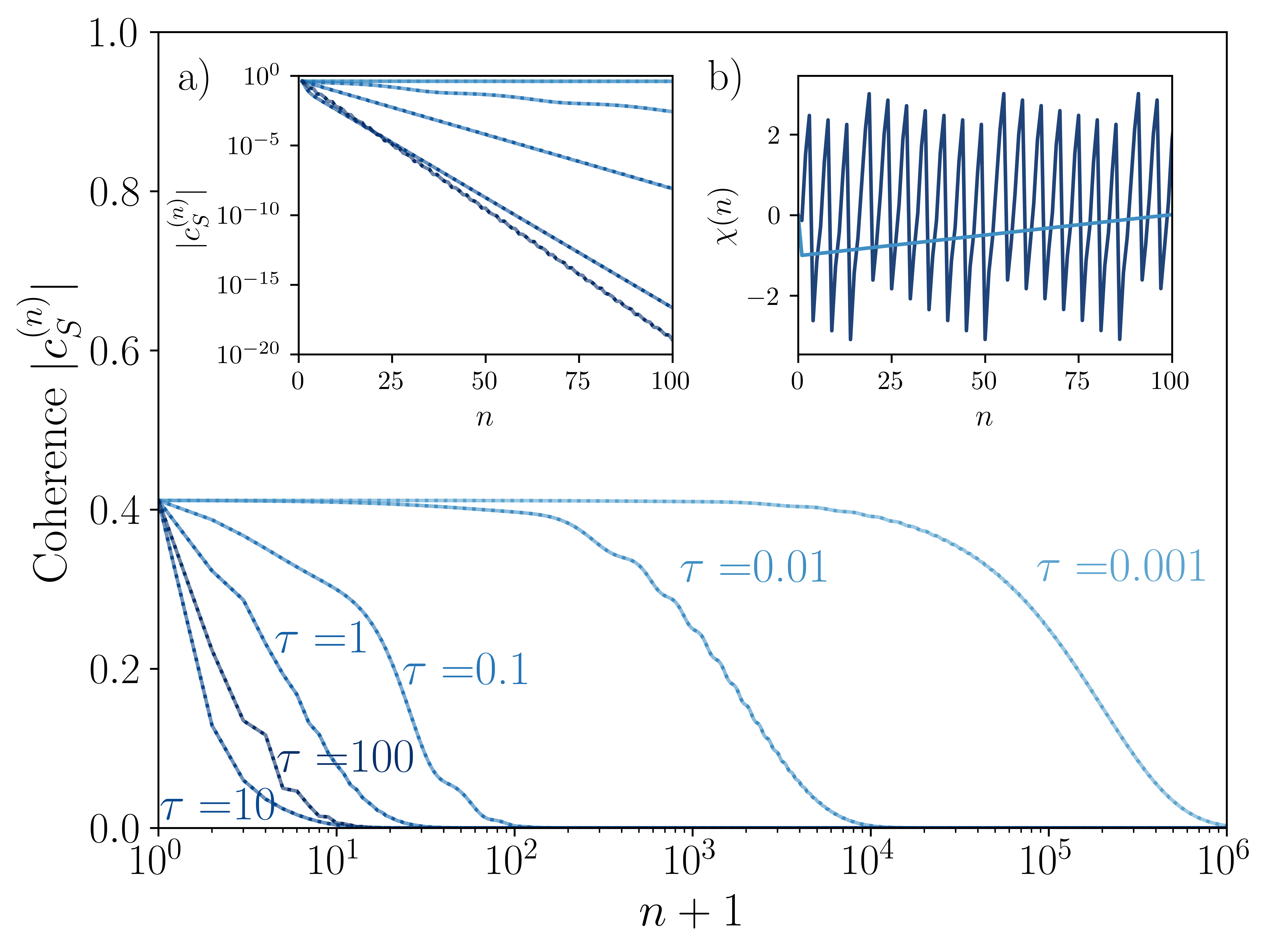} 
 \caption{Time evolution of the system coherence, $|c_S^{(n)}|$, in the  $J_{xx}$-$J_{yy}$ interaction model for different collision times ranging from $\tau=100$ (dark) to $\tau=0.001$ (light). Dynamics was simulated for $10^{6}$ steps, with $\omega_{A} = 2$, $\omega_{S} = 1$, $J_{xx} = 2$, $J_{yy}=1$. 
The initial state of the system was taken random, and the same initial state was adopted for all values of $\tau$. We plot simulation results (full) and prediction obtained  from Eqs. (\ref{eq:psidef})-\ref{eq:psi}) (dotted), overlapping.
The environmental ancillas are prepared at a Gibbs thermal state with $\beta=1$.
The main plot uses a logarithmic scale for $n$ to highlight the long-term dynamics. (a) A linear-scale representation with $n$ of the early dynamics.
(b) Display of the $n$ dependence of the phase $\chi$ for $\tau=0.01$ (light) and $\tau=100$ (dark). 
}
 \label{fig:coh1}
\end{figure}

We now return to Eq. (\ref{eq:RI}). As mentioned, simplifying this equation shows that the relaxation dynamics of population and coherences are decoupled in the model. The iterative evolution of coherences $c_{S}^{(n+1)}$ satisfies
\bea
c_{S}^{(n+1)} &=& 
(c_{S}^{(n)})^{*}
\left(J_{xx}^{2}-J_{yy}^{2}\right)
\Biggl[\frac{4}{\theta\phi}\sin \left(\frac{\theta  \tau }{2}\right) \sin \left(\frac{\phi\tau }{2}\right)p_A  + \frac{4 }{\theta  \phi }(1-p_{A})\sin \left(\frac{\theta  \tau }{2}\right)  
\sin \left(\frac{\phi\tau }{2}\right)\Biggr] 
\nonumber\\
&+& 
\Bigg[ (1-p_{A})\Biggl( \cos \left(\frac{\theta  \tau }{2}\right)-i\frac{ \left(\omega _A-\omega _S\right) }{\theta }\sin \left(\frac{\theta  \tau }{2}\right)\Biggr)
\Biggl(\cos \left(\frac{\phi\tau }{2}\right)+i\frac{ \left(\omega _A+\omega _S\right) }{\phi}\sin \left(\frac{\phi\tau }{2}\right)\Biggr) 
\nonumber\\
&+&
p_A \Biggl( \cos \left(\frac{\theta  \tau }{2}\right)-i\frac{ \left(\omega _A-\omega _S\right) }{\theta }\sin \left(\frac{\theta  \tau }{2}\right)\Biggr) \Biggl
(\cos \left(\frac{\phi\tau }{2}\right)+i\frac{ \left(\omega _A+\omega _S\right) }{\phi }\sin \left(\frac{\phi\tau }{2}\right)\Biggr)\Bigg]c_{S}^{(n)},
\label{eq:cohnp1}
\eea
and it evolves independently of populations.
We express the coherences in 
a polar form as $c_{S}^{(n)} = |c_{S}^{(n)}| e^{i\chi}$ and $(c_{S}^{(n)})^{*} = |c_{S}^{(n)}|e^{-i\chi}$.
This allows us to simplify the above equation to 
\bea
c_{S}^{(n+1)} =   \psi|c_{S}^{(n)}|
\label{eq:psidef}
\eea
 with
\begin{eqnarray}
\label{eq:psi}
  \psi &=&  \frac{4(J_{xx}^{2}-J_{yy}^{2})}{\theta\phi}e^{-i\chi}\sin\left(\frac{\theta\tau}{2}\right) \sin\left(\frac{\phi\tau}{2}\right)  
\nonumber\\
&+& 
  e^{i\chi}\left[\cos\left(\frac{\theta\tau}{2}\right)- \frac{i(\omega_{A}-\omega_{S})}{\theta}\sin\left(\frac{\theta\tau}{2}\right)\right] \cdot 
 \left[\cos\left(\frac{\phi\tau}{2}\right) + \frac{i(\omega_{A}+\omega_{S})}{\phi}\sin\left(\frac{\phi\tau}{2}\right)\right].
\end{eqnarray}
Here, the function $\psi$ dictates the evolution of coherences to their long-time solution, paralleling $\eta$ of Eq. (\ref{eq:eta2}).  The function $\psi$ displays the following properties:
(i) It generally depends on the time step $\tau$. (ii) It does not depend on the temperature of the ancilla. These two properties mirror the behavior of populations. (iii) The convergence of coherences to the null steady-state limit differs from the population's. In Eq. (\ref{eq:geom}), we show that populations decay to the steady-state value in a geometric manner. In contrast, the dynamics of coherences is more compound: At each step, the phase, obtained from $c_{S}^{(n)} = |c_{S}^{(n)}| e^{i\chi}$, may change. That is, $\chi$ varies as the system repeatedly collides with ancillas. This behavior is exemplified in Fig. \ref{fig:coh1}(b).

What is the steady-state value of coherences? If $0<|\psi|<1$, then coherences approach zero in the long-time limit. Providing a complete analytical proof for these lower and upper bounds is cumbersome. In Appendix \ref{appendB} we discuss numerical verifications and some analytical results for the limit of small collision time, $\tau \ll1$.

In the special case of $J_{xx} =J_{yy} = J$,  Eq. (\ref{eq:psi}) becomes
\begin{equation}
    \label{psiIfJxequalJy}
    \hspace{-2cm}
\psi=    
e^{i\chi}\left[\cos\left(\frac{\theta\tau}{2}\right)- i\frac{(\omega_{A}-\omega_{S})}{\theta}\sin\left(\frac{\theta\tau}{2}\right)\right] \left[\cos\left(\frac{\phi\tau}{2}\right) + i\frac{(\omega_{A}+\omega_{S})}{\phi}\sin\left(\frac{\phi\tau}{2}\right)\right].
\end{equation}
With trigonometric identities, this can be written as 
%
\begin{equation}
    \hspace{-3cm}
    \label{psiSquaredifJxequalJySimplified}
    |\psi|^{2} = \left[1-\left(1-\frac{(\omega_{A}-\omega_{S})^{2}}{\theta^{2}}\right)\sin^{2}\left(\frac{\theta\tau}{2}\right)\right]\left[1-\left(1-\frac{(\omega_{A}+\omega_{S})^{2}}{\phi^{2}}\right)\sin^{2}\left(\frac{\phi\tau}{2}\right)\right].
\end{equation}
Since in the limit $J_{xx}\rightarrow J_{yy}\equiv J$,
\begin{equation}
    \theta =  \sqrt{4(J_{xx}+J_{yy})^{2} + (\omega_{A}-\omega_{S})^2} \rightarrow\sqrt{16J^{2} + (\omega_{A}-\omega_{S})^{2}},
\end{equation}
\begin{equation}
    \phi =  \sqrt{4(J_{xx}-J_{yy})^{2} + (\omega_{A}+\omega_{S})^2}\rightarrow \omega_{A}+\omega_{S},
\end{equation}
we can then write
\begin{equation}
\label{eq:psi2J}
    |\psi|^{2} = \left[1-\frac{16J^{2}}{\theta^{2}}\sin^{2}{\left(\frac{\theta\tau}{2}\right)}\right].
\end{equation}
It is easy to show that in this case, $0\leq|\psi|^2\leq1$.

In Fig. \ref{fig:coh1}, we present the behavior of the qubit system coherences as they evolve in time from a random initial density matrix. We observe that: (i) the decay process generally depends on the timestep duration $\tau$; (ii) the phase $\chi$ change with $n$; (iii) coherences vanish in the steady-state limit.
In the main plot, we compare numerical simulations of the RI to analytical results based on Eqs. (\ref{eq:psidef})-(\ref{eq:psi}). To calculate the theoretical prediction, we computed the phase $\chi$ at each step using the relation  $c_{S}^{(n)} = |c_{S}^{(n)}| e^{i\chi}$. 
As can be shown (lines overlapping), we obtain a perfect agreement between numerical and analytical results.
Fig. \ref{fig:coh1} (a) also shows that, while the decay of coherences is close to an exponential in $n$, it is not exactly so, as manifested by some oscillations in the coherences. Furthermore, in Fig. \ref{fig:coh1}(b) we observe a periodic trend for the phase, with the period becoming shorter when increasing $\tau$.

 \begin{figure}[htbp]
    \centering
    \includegraphics[width=0.75\linewidth]{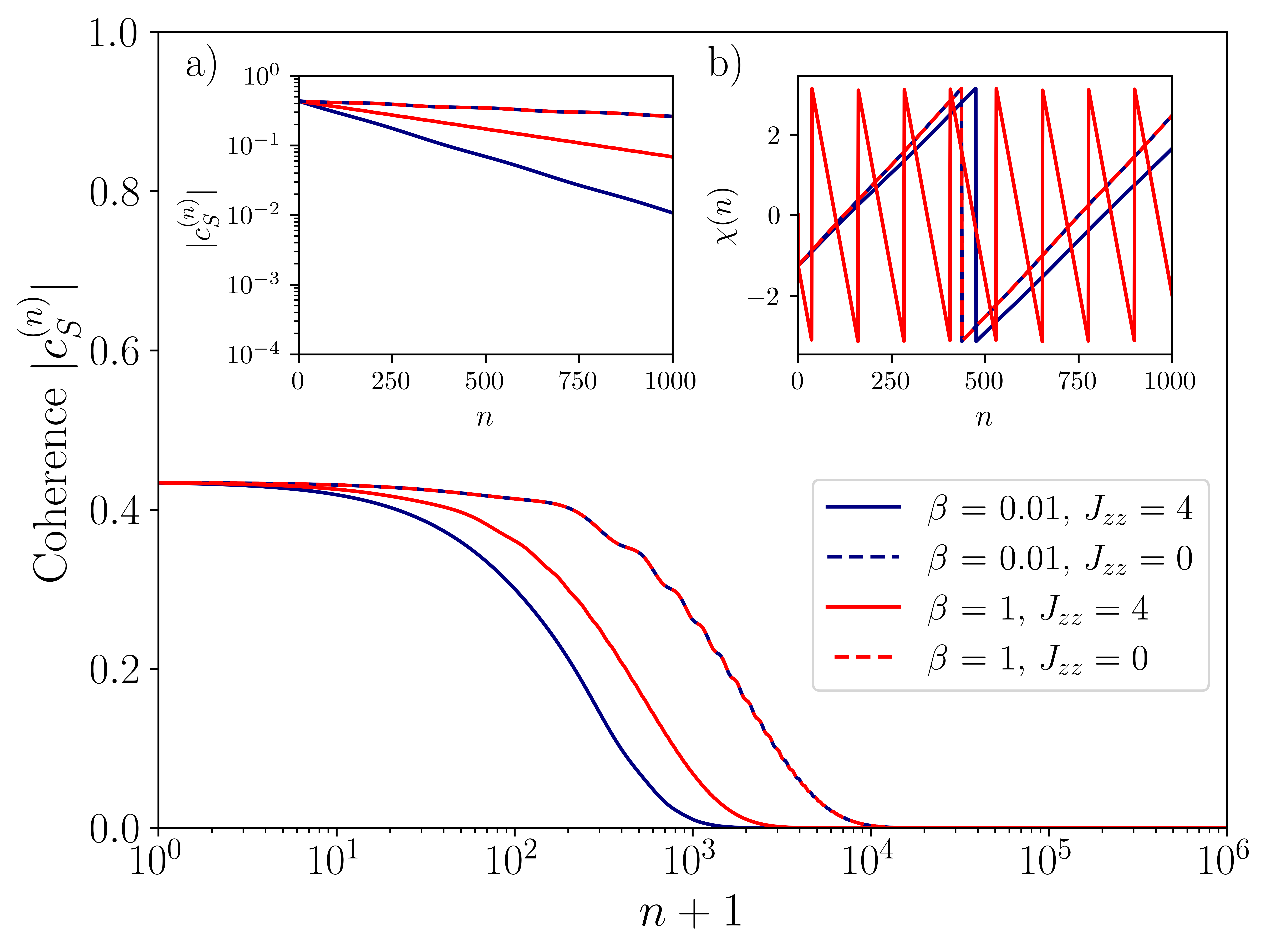}
\caption{Time evolution of the system coherence, $|c_{S}^{(n)}|$, for different values of the inverse temperature of the bath, namely $\beta = 0.01$ and $\beta = 1$, for $J_{zz} = 0$ (dashed, overlapping) and $J_{zz}=4$ (full). 
Parameters are $J_{xx}=2$, $J_{yy} = 1$, $\omega_{A}=2$, $\omega_{S}=1$. The collision time is fixed at $\tau=0.01$. 
The initial state of the system is random with non-vanishing coherences, and the same state was adopted in all simulations. The state of each ancilla before collision is a Gibbs thermal state with the inverse temperature $\beta$.}
\label{fig:beta} 
\end{figure}


\section{RI dynamics and steady state: $J_{xx}-J_{yy}-J_{zz}$ model}
\label{sec:XYZ}

\subsection{Model and simulations}

The analysis presented in Sec. \ref{sec:XY} can be generalized to describe a Heisenberg-type interaction Hamiltonian between the system $S$ and the environmental ancilla $A$,
\begin{equation}
\label{HSHAZ}
    \hat{H}_{I} = J_{xx}\hat{\sigma}_{x}^{S}\otimes \hat{\sigma}_{x}^{A} +J_{yy}\hat{\sigma}_{y}^{S}\otimes \hat{\sigma}_{y}^{A}
    +J_{zz}\hat{\sigma}_{z}^{S}\otimes \hat{\sigma}_{z}^{A}.
\end{equation}
Here, $\hat\sigma_{z}^{\bullet}$ is the $z-component$ of the Pauli spin operators and $J_{zz}$ represents the corresponding coupling strength.
In matrix representation, the total Hamiltonian at each iteration, $\hat H_{\text{tot}}$, becomes
\begin{equation}
\hat{H}_{\text{tot}} =
\begin{pmatrix}
 J_{zz}-\frac{\omega _A}{2}-\frac{\omega _S}{2} & 0 & 0 & J_{xx}-J_{yy} \\[0.25cm] 
 0 & -J_{zz}+\frac{\omega _A}{2}-\frac{\omega _S}{2} & J_{xx}+J_{yy} & 0 \\[0.25cm] 
 0 & J_{xx}+J_{yy} & -J_{zz}-\frac{\omega _A}{2}+\frac{\omega _S}{2} & 0 \\[0.25cm] 
 J_{xx}-J_{yy} & 0 & 0 & J_{zz}+\frac{\omega _A}{2}+\frac{\omega _S}{2} \\[0.25cm] 
\end{pmatrix}
\end{equation}
The collision unitary can be explicitly computed, similarly to Eq. (\ref{eq:U}) revealing a natural decoupling of the dynamics into two distinct subspaces: one associated with a single spin excitation (centre block) and the other encompassing zero or two excitations (outer envelope). 
Furthermore, when analyzing the RI dynamics under Eq. (\ref{HSHAZ}), just as with Eq. (\ref{eq:HI_JxxJyy}), we find that the populations and coherences evolve independently and each component asymptotically approaches its steady-state value without mutual influence.
From the analytical procedure detailed in Sec. \ref{sec:XY}, the steady state of the system assumes once again a diagonal form {\it identical} to that obtained for the interaction Hamiltonian in Eq. (\ref{eq:HI_JxxJyy}). Specifically, the ground-state population $p_{S}^{(\infty)}$ is given by the analytical expression  reported in Eq. (\ref{eq:psss}).
With regard to coherences, numerical results confirm that they decay to zero. However, their time evolution now exhibits a dependence on the ground-state population of the environmental ancilla, $p_A$, which, through Eq. (\ref{AncillaInitialState}), is determined by the temperature of the ancilla.
In particular, we obtain for the coherences of the system  
\begin{equation}
c_{S,z}^{(n+1)} = \Tilde{\psi}|c_{S,z}^{(n)}|,    
\label{eq:coh_Jzz}
\end{equation}
where we highlight with the subscript $z$ the role of the $J_{zz}$ interaction.
The function $\Tilde{\psi}$ is given by
\bea    
\Tilde \psi&=&
\frac{4(J_{xx}^{2}-J_{yy}^{2})}{\theta\phi}e^{-i(\chi+2J_{zz}\tau)}\sin\left(\frac{\theta\tau}{2}\right)\sin\left(\frac{\phi\tau}{2}\right) 
\nonumber\\
&+&
     e^{i(\chi+2J_{zz}\tau)}\left(\cos\left(\frac{\theta\tau}{2}\right)-i\frac{\left(\omega_{A}-\omega_{S}\right)}{\theta}\sin\left(\frac{\theta\tau}{2}\right)\right)\left(\cos\left(\frac{\phi\tau}{2}\right)+i\frac{\left(\omega_{A}+\omega_{S}\right)}{\phi}\sin\left(\frac{\phi\tau}{2}\right)\right) 
\nonumber\\
&+&
    2ip_{A}\Biggl[\frac{4(J_{xx}^{2}-J_{yy}^{2})}{\theta\phi}e^{-i\chi}\sin\left(\frac{\theta\tau}{2}\right) \sin\left(\frac{\phi\tau}{2}\right)  - 
    e^{i\chi}\left(\cos\left(\frac{\theta\tau}{2}\right)- \frac{i(\omega_{A}-\omega_{S})}{\theta}\sin\left(\frac{\theta\tau}{2}\right)\right)\cdot 
\nonumber\\
& \cdot&\left(\cos\left(\frac{\phi\tau}{2}\right) + \frac{i(\omega_{A}+\omega_{S})}{\phi}\sin\left(\frac{\phi\tau}{2}\right)\right) \Biggr]\sin(2J_{zz}\tau),
\label{eq:psitilde}
\eea
where $\theta$ and $\phi$ are the energy parameters given by Eq. (\ref{eq:thetaphi}). Similarly to what we found in Sec. \ref{sec:XY}, is it difficult to prove that $|\Tilde{\psi}|<1$. Once again, we verify this bound numerically as described in Appendix \ref{appendA}, and with $J_{zz}$ randomly selected from a uniform distribution $U(-100,100)$.

We conclude that the additional interaction term $\propto \hat{\sigma}_{z}^{S} \otimes \hat{\sigma}_{z}^{A}$ in the RI Hamiltonian impacts only coherences and not population. It does so in two ways. First, it introduces a dependence of the coherence dynamics on the ancilla temperature through a term proportional to $p_{A}$. Second, it contributes an additional phase factor to the time evolution of coherences, as previously noted in Ref. \citenum{Campbell20}.

In Fig. \ref{fig:beta}, we illustrate the temperature dependence of coherence dynamics due to the $J_{zz}$ coupling: When $J_{zz}=0$, the decay is independent of temperature (dashed).  In contrast, once $J_{zz}\neq 0$, we find that the decay is faster when the temperature is raised (full), in line with general expectations. We also show in Fig. \ref{fig:beta}(a) that the decay is approximately exponential with $n$, and that the phase varies with temperature once $J_{zz}\neq 0$, see Fig. \ref{fig:beta}(b).



\subsection{The steady state as a global fixed point for the dynamical map}

We summarize Secs. \ref{sec:XY}-\ref{sec:XYZ} by pointing out the following. 
The steady-state reached by the system $S$ under the dynamics generated by the Heisenberg-type Hamiltonians represents a \emph{global fixed point} for the corresponding dynamical maps. 
We start by recalling the definition of global fixed point of a map.
\begin{definition}
    Let $\mathcal{H}$ be the Hilbert space of a quantum system, and $\mathcal{D}(\mathcal{H})$ the space of positive, trace class linear operators acting on $\mathcal{H}$, as density matrix operators. A quantum map $\mathcal{E}[\rho]: \mathcal{D}(\mathcal{H})\rightarrow\mathcal{D}(\mathcal{H})$ is said to \emph{admit a global fixed point} if there exists a unique
    $\rho^{*}$ such that $\:\:\mathcal{E}[\rho^{*}] = \rho^{*}$.
\end{definition}
Looking at dynamical maps of previous sections, obtained using the ansatz of Eq. (\ref{eq:ansatz}), we note that the steady state $\rho_{S}^{(\infty)}$ is  {\it independent} of the initial state of the reduced system $\rho_{S}^{(0)}$; it depends 
only on the model parameters and $\rho_A$.
Obviously, the model parameters also set the dynamical map that governs the time evolution of the state of the system, fixing $\eta, \psi$ (or $\Tilde{\psi}$).
Except for some special combinations of parameters giving Rabi oscillations-like behavior (namely $\eta=1$, see Appendix \ref{appendA}), the dynamics of the system generated by the maps derived in Secs. \ref{sec:XY}-\ref{sec:XYZ} always converge to its unique fixed point represented by $\rho_{S}^{(\infty)}$.


\section{Thermalization with a few long randomized RI steps}
\label{sec:random}

Our main result, the relaxation equation (\ref{eq:ansatz}) with the rate (\ref{eq:eta2}) and fixed point for population (\ref{eq:psss}) provides the relaxation dynamics of population to the steady state limit. In general, 
 if interaction coupling strengths are different, i.e., $J_{xx}\neq J_{yy}$, the system does not thermalize to the state of the ancillas.
Focusing on Eq. (\ref{eq:psss}), only when $J_{xx}=J_{yy}$, the state of the system thermalizes, with the ground state reaching $p_S^{(\infty)} =  p_A$, with coherences decaying to zero; note that when $\omega_A=\omega_S$ the eventual temperature of the system is different than $\beta$, but we still refer to this case as 
``thermalization" since the state of the system reaches the ancilla's \cite{commentT}. 

The main limit examined in the literature is that of {\it short} interaction time intervals.
For example, in Fig. \ref{figP}, we reach the steady state within $n > 100$ once $\tau\ll1$ with $J$ order of 1, that is, $J\tau\ll1$. Here, $J$ stands as a general notation for the coupling energy.

However, there is a {\it special} limit to Eq. (\ref{eq:psss}), where {\it thermalization can be reached approximately}. This protocol uses {\it long} interaction time intervals and very small $J$ terms, as we describe next. Consider an RI model satisfying the following requirements:

(i) The system qubit and the ancilla share the same frequency, $\omega=\omega_S=\omega_A$.

(ii) Both $J_{xx}$ and $J_{yy}$ are small, 
$J_{xx}, J_{yy}\ll\omega$. 
That is, the interaction energy is {\it small} relative to the bare system and ancilla's frequencies. Note that we do not enforce the interactions $J_{xx}$ and $J_{yy}$ to be identical. In fact, they can be randomly chosen at every RI step, as long as they satisfy the condition that they are small.

(iii) $J \tau$ is of order 1, with $J$ a characteristic value for $J_{xx}$ and $J_{yy}$. However, since we imposed in (ii) that $J$ is small relative to $\omega$, the present condition means that the interaction time should be made {\it long}, unlike the limits we considered in Secs. \ref{sec:XY} and \ref{sec:XYZ}.

These three requirements lead to 
$\phi\to 2\omega$, $\theta\to 4J$, $\phi\gg \theta $, 
and to $\theta \tau$ being of order 1, see Eq. (\ref{eq:thetaphi}). That is, we {\it cannot} generally assume that $\sin(\theta\tau)\sim \theta \tau$ or
$\sin(\phi\tau)\sim \phi \tau$.

Let us begin by inspecting $\eta$ in Eq. (\ref{eq:eta2}). We focus on the two fractions and note that under our setup, 
\bea
\frac{ (J_{xx}+J_{yy})^2}{\theta^2} \sin^2(\theta \tau/2)\lesssim 1
\nonumber\\
\frac{ (J_{xx}-J_{yy})^2}{\phi^2} \sin^2(\phi \tau/2) \ll1
\eea
As a result, $\eta\ll1$. The qubit system will thus {\it quickly}, in a few RI iterations, reach the steady state. What is the steady state? We simplify Eq. (\ref{eq:psss}) by identifying dominant contributions. We find that
%
\bea
&&p_S^{(\infty)}  \xrightarrow{\theta\ll\phi, \theta\tau = O(1)}  p_A. 
\eea
Thus, the populations of the system approach that of the ancilla. 
As for coherences, we analyze Eq. (\ref{eq:psi}). The second term is small in our limit of $\theta\sim4J$, see Eq. (\ref{eq:psi2J}).
The first term in Eq. (\ref{eq:psi}) is small given that $(J_{xx}^2-J_{yy}^2)\ll \theta\phi$. As a result, $|\psi|^2\ll1$.
Similarly to the behavior of populations, coherences decay within few steps to
zero. Alltogether, in the suggested protocol, the system approaches its fixed point in a few long RI collisions. The fixed point reached by the system corresponds to the ancilla's thermal state.

\begin{figure}[htbp]
\includegraphics[width=0.75\linewidth]{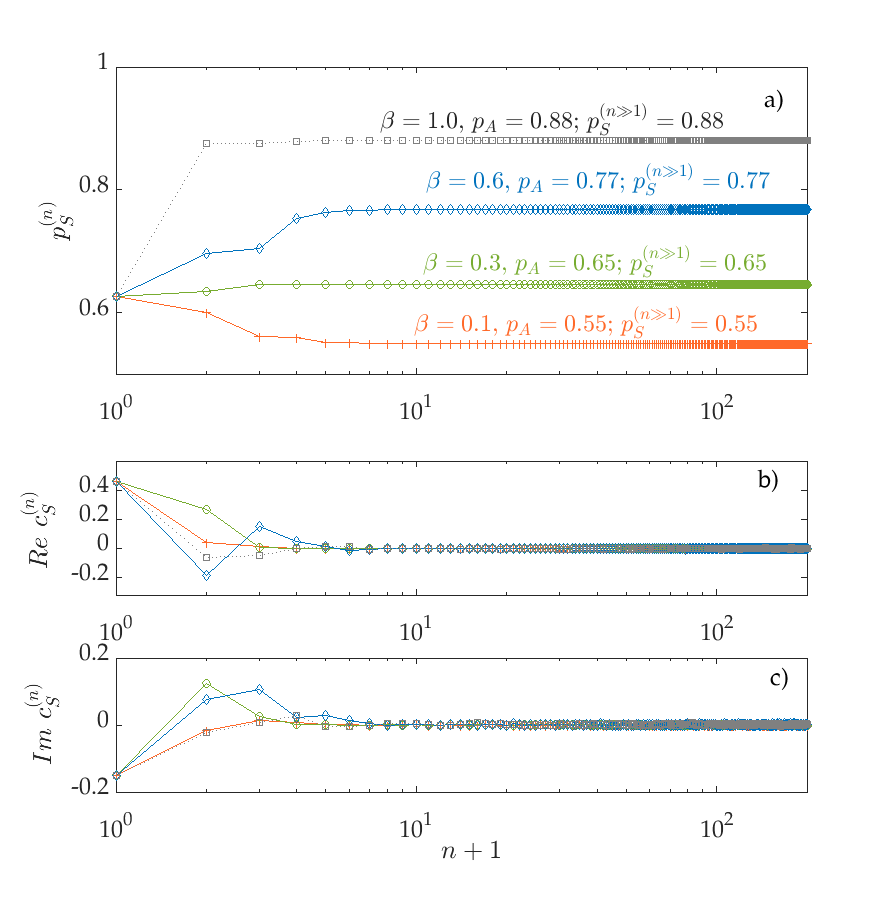} 
\caption{Thermalization process in the $J_{xx}-J_{yy}$ model with a few long RI collisions. 
We present the evolution of (a) ground state population ($p_S^{(n)}$), (b) real  ($Re\,c_S^{(n)}$) and (c) imaginary ($Im \,c_S^{(n)}$) parts of the state of the system from the initial condition at $n=0$.
Parameters are $\omega_A=\omega_S=2$;
$J_{xx}$ and $J_{yy}$ are sampled at each iteration from a uniform distribution in the range $[0, 0.01]$, $\tau=100$.
The initial conditions included coherences, as identified by the first data point.
 }
 \label{fig:fast}
\end{figure}

Fig. \ref{fig:fast} exemplifies this thermalization dynamics. We use $\omega_S=\omega_A$, and sample the system-ancilla interaction energies $J_{xx}$ and $J_{yy}$ from a uniform distribution in the range $[0, 0.01]$. 
Importantly, the collision time $\tau$ is taken large enough such that $J\tau$ is order of 1.
It is clear that the system thermalizes within about five RI steps. It is also understandable, however, that the states of the ancillas that are interacting within these first timesteps are being largely perturbed from their initial thermal state after a collision. To make this process useful, it is interesting to assess this RI thermalization protocol on a multilevel system, ot demonstrate thermalization of a generic system by a collection of ancillas.

Regarding the absolute time required by the system for reaching the steady state, namely to thermalize, in Fig. \ref{figP} the time scale adopted corresponds to $\tau n \approx 100 $ with, for instance, $\tau=0.01$ and $n=10^4$. The thermalization reported in Fig. \ref{fig:fast} is characterized by a collision time $\tau=100$ and it occurs with $n$ up to 5. Thus, the ``real" time for reaching steady state is similar whether we do it with short and rapid steps (reaching a nonequilibrium steady state), or with long and a few steps (achieving thermalization). 

We described here an RI scheme with {\it random} but small coupling parameters, which achieved thermalization.
While we demonstrated here results for the $J_{xx}-J_{yy}$ model, simulations carried out with more general interaction Hamiltonians lead to the same observation of a few-step thermalization. 

In the rest of the paper, we turn to resource estimation, but return to the short-time RI problem, which we considered in Secs. \ref{sec:XY}-\ref{sec:XYZ}. We focus on short collision times for two reasons. First, this is the limit that is appropriate for deriving the Lindblad, or other dynamical equations of motion. Second, foreseeing implementations on a quantum hardware, long gate operations are not favorable as they will suffer more errors. Nevertheless, thermalization under a few long randomized RI steps is a topic of a great interest, to be explored in future studies. 

\section{Resource estimation: Number of RI iterations and energetic cost}
\label{sec:energetics}

In this section, we analyze the resources associated with the RI process carried out to {\it create} the nonequilibrium steady state. First, we determine the number of collisions---or iterations of the RI protocol---that are required to bring the system sufficiently close to its steady-state value. The number of iterations can be mapped to the {\it runtime} of the algorithm, or the required number of gates, when the RI scheme is implemented as a quantum circuit.
Next, we estimate the {\it energy} cost of the process by evaluating the amount of work associated with turning on and off the system-ancilla interaction until creating the steady state within the prescribed accuracy.


\subsection{Number of RI steps/runtime}

The distance between the system's state at a given time step and its steady state value can be quantified using various metrics. Here, we employ two commonly used measures: trace distance and fidelity. Using the trace distance, we are able to lower-bound the number of RI steps required to approach the steady state within a certain accuracy, so long as the state of the system is diagonal to begin with.
For fidelity, we provide working expressions without simplifying them.

\subsubsection{Trace Distance}
\label{sec:T}

The \textit{Trace Distance} 
between two matrices, $\rho$ and $\sigma$, within the same vector space, is defined as \cite{Nielsen}
\begin{equation}
   \label{eq:TraceDistanceDef}
   ||\rho-\sigma||_{\rm Tr} 
= \frac{1}{2}{\rm Tr}\left(\sqrt{(\rho-\sigma)(\rho-\sigma)^{\dagger}}\right).
\end{equation}
For the models under investigation, the quantity of interest is $||\rho_{S}^{(n)}-\rho_{S}^{(\infty)}||_{\rm Tr}$. Observing that
\begin{equation}
\small
    \hspace{-4cm}
    \label{TraceDistanceOurCase}
    (\rho_{S}^{(n)}-\rho_{S}^{(\infty)})(\rho_{S}^{(n)}-\rho_{S}^{(\infty)})^{\dagger} = 
     \begin{pmatrix}
         p_{S}^{(n)} - p_{S}^{(\infty)}  & c_{S}^{(n)} \\
         c_{S}^{(n)*} &  p_{S}^{(\infty)}-p_{S}^{(n)} 
     \end{pmatrix}^2
     %
\end{equation}
it is possible to write the trace distance definition as 
\bea    
||\rho_{S}^{(n)}-\rho_{S}^{(\infty)})||_{\rm Tr} 
&=&  \sqrt{ (p_{S}^{(n)})^{2} + (p_{S}^{(\infty)})^{2} - 2p_{S}^{(n)}p_{S}^{(\infty)} + |c_{S}^{(n)}|^{2}} 
\nonumber\\
&=&  \sqrt{ \left( p_{S}^{(n)} - p_{S}^{(\infty)}\right)^{2} + |c_{S}^{(n)}|^{2}}  
\nonumber\\
    &=&  \sqrt{  \eta^{2n} \left(p_{S}^{(0)} - p_{S}^{(\infty)}\right)^{2} 
+ \Pi_n|\psi(n)|^2|c_{S}^{(0)}|^{2}}.
\eea
In the last line, we used the solution (\ref{eq:geom}) for the ground-state population and the expression for coherences reported in Eq. (\ref{eq:psidef}) or (\ref{eq:coh_Jzz}), depending on the system-ancilla interaction model. It is important to remember the dependence of the coefficients $\psi$ (for the $J_{xx}-J_{yy}$ model) or $\tilde \psi$ (for the $J_{xx}-J_{yy}-J_{zz}$ model) on the iteration step, $n$, through the phase factor $\chi$.

We aim to compute the number of iterations, $n^{*}$, necessary to satisfy   
\bea
   ||\rho_{S}^{(n^{*})}-\rho_{S}^{(\infty)})||_{\rm Tr} \leq  \epsilon,
   \label{eq:traceeps}
\eea
in correspondence of some threshold $\epsilon\in\mathbb{R}_{+}$. We thus have the following equation for $n^{*}$, 
\bea
\sqrt{ \eta^{2n^*}\left( p_{S}^{(0)} - p_{S}^{(\infty)}\right)^{2}
+ \Pi_{n^*}|\psi(n^*)|^2|c_{S}^{(0)}|^{2}} \leq \epsilon.
\label{eq:problem}
\eea
We can find $n^*$ numerically for any initial conditions.  Furthermore, Eq. (\ref{eq:problem}) can be simplified if we limit ourselves to a diagonal initial state for the density matrix of the system,
since we found that no coherences will be generated on the go with our choices for interaction Hamiltonians.
Taking $|c_{S}^{(0)}| = 0$, this equation simplifies to
$     |\eta|^{n^*} \left|p_{S}^{(0)}-p_{S}^{(\infty)}\right| \leq \epsilon$,
 namely,
\begin{equation}
n^* \geq \ln\left(\frac{\epsilon}{\left|p_{S}^{(0)}-p_{S}^{(\infty)}\right|}\right) \times \frac{1}{\ln(|\eta|)}.
\label{eq:nstar}
\end{equation}
This result was derived by assuming $0<|\eta|<1$. That is, the relaxation rate $\eta$ should not hit the specific set of points where the upper limit is reached; see Appendix \ref{appendA}. In addition, the collision time $\tau$ is assumed to be finite and non-zero, therefore, we do not hit the lower bound of zero for $\eta$. 

Equation (\ref{eq:nstar}) is one of our main results. It indicates that to reduce the system's distance from a diagonal initial condition to within $\epsilon$ from its steady state, at least $\ln (C\epsilon)/\ln |\eta|$ RI steps are required, where $C>1$ is a constant determined by the distance of initial conditions from the final state. As expected, when $\eta\to 1$, many iterations are needed to approach the steady state, whereas a smaller $\eta$ corresponds to a faster relaxation process.
Since for short collisions, $\eta\approx 1-4J^2\tau^2$, see Eq. (\ref{eq:ptauss}), $\ln \eta \approx -4 J^2\tau^2$ and $n^* \propto(J\tau)^{-2}$.
As a reminder, Eq.  (\ref{eq:nstar}) holds for both interaction models considered here, regardless of whether $J_{zz}$ is nonzero.

\subsubsection{Fidelity}
\label{sec:F}

The \textit{Fidelity} $F(\rho,\sigma)$ between two matrices, $\rho$ and $\sigma$, within the same vector space, is defined according to \cite{Nielsen}  
\begin{equation}
\label{eq:FidelityD}
    F(\rho,\sigma) = \left({\rm Tr}\left[\sqrt{\sqrt{\rho}\sigma\sqrt{\rho}}\right] \right)^{2}.
\end{equation}
If the state of the system after  $n$ steps is diagonal, namely, $\rho_{S}^{(n)} = diag(p_{S}^{(n)},1-p_{S}^{(n)})$, and the steady state of the system, $\rho_{S}^{(\infty)}$ is diagonal, then the definition of fidelity, given in Eq. (\ref{eq:FidelityD}) can be rewritten as  
\bea    
F(\rho_{S}^{(n)},\rho_{S}^{(\infty)}) &=& \left({\rm Tr}\left(\sqrt{\sqrt{\rho_{S}^{(n)}}\rho_{S}^{(\infty)}\sqrt{\rho_{S}^{(n)}} }\right)\right)^{2} 
\nonumber\\
    &=& \left({\rm Tr}\left(\sqrt{diag\left(\sqrt{p_{S}^{(n)}},\sqrt{1-p_{S}^{(n)}}\right)diag\left(p_{S}^{(\infty)},1-p_{S}^{(\infty)}\right)diag\left(\sqrt{p_{S}^{(n)}},\sqrt{1-p_{S}^{(n)}}\right)}\right)\right)^{2} 
\nonumber\\
    &=& \left(\sqrt{p_{S}^{(n)}p_{S}^{(\infty)}} + \sqrt{(1-p_{S}^{(n)})(1-p_{S}^{(\infty)})}\right)^{2} 
\nonumber\\
    &=& 1-2p_{S}^{(\infty)}+2(p_{S}^{(\infty)})^{2}+ \eta^{n}\left(p_{S}^{(0)}-p_{S}^{(\infty)}\right) \left(2 p_{S}^{(\infty)}-1\right) 
\nonumber\\
    &+&  2 \sqrt{\left(p_{S}^{(\infty)}-1\right) \left(\left(p_{S}^{(0)}-p_{S}^{(\infty)}\right) \eta ^n+p_{S}^{(\infty)}-1\right)} \sqrt{p_S^{(\infty)} \left(\left(p_{S}^{(0)}-p_{S}^{(\infty)}\right) \eta ^n+p_{S}^{(\infty)}\right)},
    \label{eq:Fidelity2}
\eea
where we used Eq. (\ref{eq:ansatz}) and the steady state solution $\rho_{S}^{(\infty)} = diag(p_{S}^{(\infty)},1-p_{S}^{(\infty)})$, as derived in Secs. \ref{sec:XY} and \ref{sec:XYZ}. 
Recall that the time-evolving state of the system is diagonal if the initial state of the system is diagonal, the environmental ancilla is in a diagonal, and the interaction Hamiltonian is of the Heisenberg form of Eq. (\ref{eq:HI_JxxJyy}) or Eq. (\ref{HSHAZ}).

To find $n^*$, we need to simplify the inequality
\bea
1- F(\rho_{S}^{(n^*)},\rho_{S}^{(\infty)})\leq \epsilon,
\label{eq:fidelityeps}
\eea
for some threshold $\epsilon\in\mathbb{R}^{+}$. 
That is, we need the {\it infidelity} to be smaller than the threshold.
Simplifying expressions furthermore is cumbersome, even in the case of diagonal states. Therefore, unlike the trace distance, for fidelity, we calculate the number of iterations $n^*$ numerically.

\subsection{Energetic cost}\label{sec:inputwork}

Since $\hat H_{\text{tot}}$ and $\hat{U}(\tau)$ commute, at each time step the change in the expectation value of the total energy is zero,
\bea
\Delta E_{\text{tot}}^{(n+1)} = {\rm Tr} \left[ \left(U^{\dagger}(\tau)\hat H_{\text{tot}} 
\hat U(\tau)-\hat H_{\text{tot}}\right)\rho_S^{(n)}\otimes \rho_A  \right] = 0.
\label{eq:Etot}
\eea
The trace operation $\rm Tr$ stands for the total trace.
This energy change can be expressed as a statement for the first law of thermodynamics at each step,
$\Delta E_{\text{tot}} = W_I + Q+\Delta E_S=0$.
The work term arises due to the switching on and off of the interaction energy \cite{Barra},
\begin{eqnarray}
    \label{eq:workstep}
    W_I^{(n+1)} = {\rm Tr}\left[\left(\hat{U}^{\dagger}(\tau) \hat{H}_{I}\hat{U}(\tau) - \hat{H}_{I}\right)\rho_{S}^{(n)}\otimes\rho_{A}\right].
\end{eqnarray}
The heat is the amount of energy deposited to the ancillas from the system,
%
\begin{eqnarray}
    \label{eq:Qstep}
    Q^{(n+1)} = {\rm Tr}\left[\left(\hat{U}^{\dagger}(\tau) \hat{H}_{A}\hat{U}(\tau) - \hat{H}_{A}\right)\rho_{S}^{(n)}\otimes\rho_{A}\right].
\end{eqnarray}
The last contribution to total energy is related to the change in the internal energy of the system, written here as 
\begin{eqnarray}
    \label{eq:Esstep}
    \Delta E_S^{(n+1)} ={\rm Tr}\left[\hat H_S \left( \hat U(\tau) 
    \rho_{S}^{(n)}\otimes\rho_{A} \hat U^{\dagger}(\tau) 
      -\rho_{S}^{(n)}\otimes\rho_{A} \right)\right].
\end{eqnarray}
%
In this form, it is clear that once the dynamics reaches the steady-state limit, $\Delta E_S^{(\infty)}=0$ and $W_I^{(\infty)} = -Q^{(\infty)}$.
In words, in steady state, the work added to the interaction energy is wasted as heat emitted to the ancillas' bath. 
These work and heat necessary to {\it maintain} the nonequilibrium steady state are regarded as \textit{housekeeping} work and heat \cite{Campbell20}.

We next focus on the following question:
How much work is involved in the task of {\it creating}, within a certain accuracy, the nonequilibrium steady state in the RI scheme?  
The work required to perform a single RI step, as defined in Eq. (\ref{eq:RI}), is given by the change in the energy stored in the system-ancilla interaction \cite{Campbell20}. For the step $n+1$, the energetic cost is given by Eq. (\ref{eq:workstep}).
Note that this energetic cost does not include the energy needed to prepare or refresh the ancilla. 
The total work required to complete $n^{*}$ RI iterations to bring the system $\epsilon$ close to the target state, is 
\bea
W(n^*)= -\sum_{n=1}^{n^*} W_I^{(n)}.
\label{eq:totalwork}
\eea
If $W_I$ is negative, it means that the interaction energy was {\it reduced} during an RI step. We thus define the total work that should be {\it invested} to run the RI scheme up to $n^*$ by including a minus sign in front of Eq. (\ref{eq:totalwork}):
To perform a new system-ancilla collision, we need to switch on again the interaction, thus we must provide the interaction energy lost in the previous collision. This process requires performing work.

Simplifying Eq. (\ref{eq:workstep}) for an arbitrary initial state of the system, we obtain 
\begin{equation}
   W_I^{(n+1)} = 4\left[\frac{(J_{xx}+J_{yy})^{2}(\omega_{A}-\omega_{S})}{\theta^{2}}\sin^{2}{\left(\frac{\theta\tau}{2}\right)}\left(p_{S}^{(n)}-p_{A}\right)-\frac{(J_{xx}-J_{yy})^{2}(\omega_{A}+\omega_{S})}{\phi^{2}}\sin^{2}{\left(\frac{\phi\tau}{2}\right)}\left(p_{S}^{(n)}-(1-p_{A})\right)\right],
    \label{eq:workstep_analytical}
\end{equation}
A similar simplification of Eq. (\ref{eq:Qstep}) leads to
\begin{equation}
    Q^{(n+1)} = 4\omega_{A}\left[-\frac{(J_{xx}+J_{yy})^{2}}{\theta^{2}}\sin^{2}\left(\frac{\theta\tau}{2}\right)(p_{S}^{(n)}-p_{A}) +\frac{(J_{xx}-J_{yy})^{2}}{\phi^{2}}\sin^{2}\left(\frac{\phi\tau}{2}\right)\left(p_{S}^{(n)}-(1-p_{A})\right)\right].
    \label{eq:heatstep_analytical}
\end{equation}
As for the change of energy of the system, from Eq. (\ref{eq:Esstep}) we get
\begin{equation}
    \Delta E_{S}^{(n+1)} = 4\omega_{S}\left[\frac{(J_{xx}+J_{yy})^{2}}{\theta^{2}}\sin^{2}\left(\frac{\theta\tau}{2}\right)(p_{S}^{(n)}-p_{A}) +\frac{(J_{xx}-J_{yy})^{2}}{\phi^{2}}\sin^{2}\left(\frac{\phi\tau}{2}\right)\left(p_{S}^{(n)}-(1-p_{A})\right)\right].
\label{eq:energystep_analytical}
\end{equation}
Equations (\ref{eq:workstep_analytical})-(\ref{eq:energystep_analytical}) constitute another main result of this work, and we can now draw several physical observations.
First, it is easy to confirm that $W_I^{(n+1)}+Q^{(n+1)}+\Delta E_{S}^{(n+1)} = 0$.
Second, remarkably, these expressions immediately show that erasing coherence does not incur an energetic cost.
Third, we now understand work input/output in the system. Consider for simplicity the case of equal frequencies, $\omega_S=\omega_A$. Focusing on Eq. (\ref{eq:workstep_analytical}), it reveals that $W_I^{(n)}\propto -\left(p_{S}^{(n)}-(1-p_{A})\right)$. Thus, this population imbalance dictates whether work should be invested (negative sign) or gained (positive) in the process of relaxation to steady state, so long as $J_{xx}\neq J_{yy}$.
Finally, if $J_{xx}=J_{yy}$, then $p_S^{(\infty)}\to p_A$ according to 
Eq. (\ref{eq:psss}) and there is no housekeeping work required to maintain the steady state, $W_I^{(\infty)}=0$ --- regardless of the values of $\omega_{A,S}$--- implying a thermal equilibrium solution  \cite{commentT}. 

\begin{figure}
    \centering
    \includegraphics[width=0.7\linewidth]{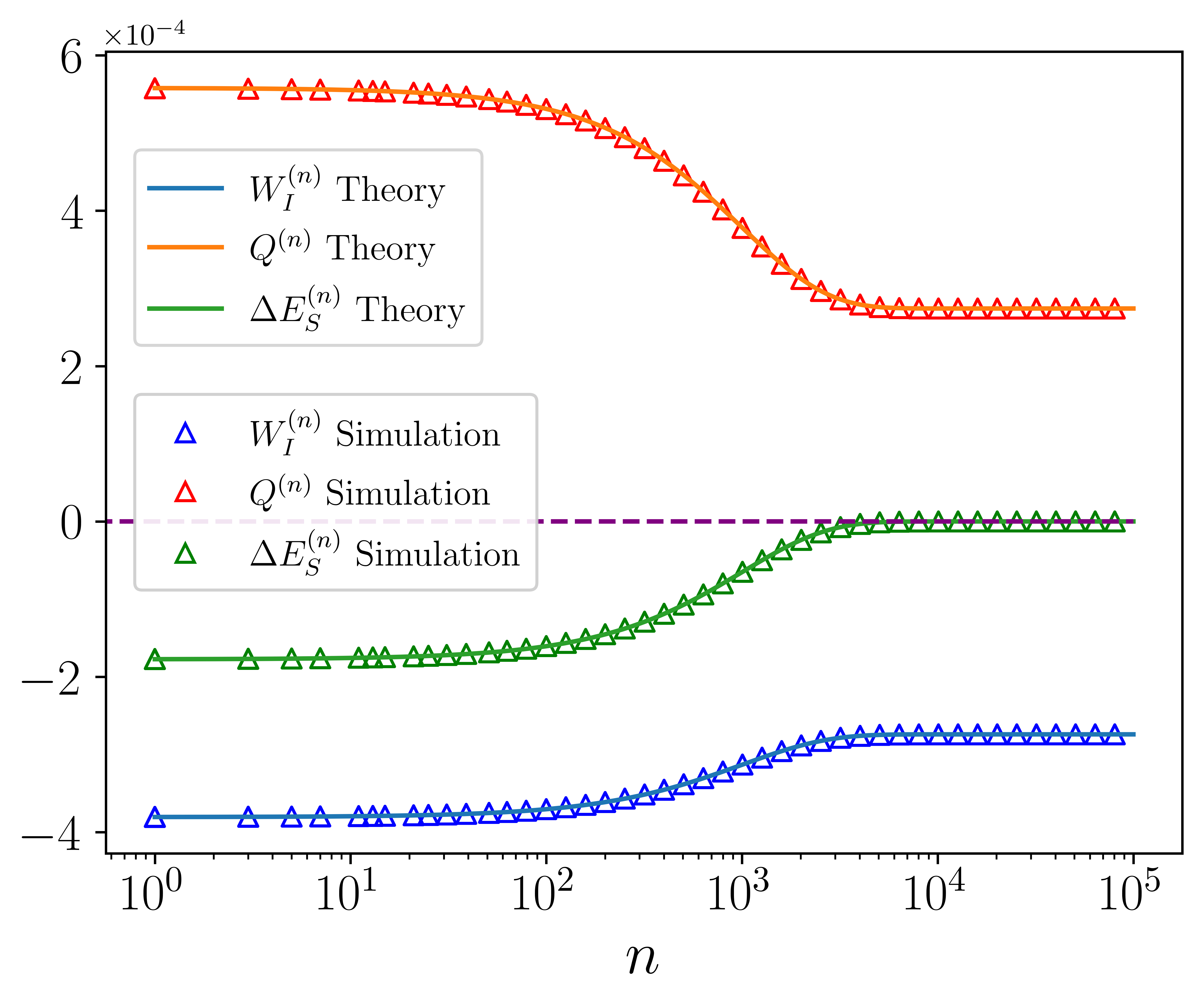} 
 \caption{Work, heat and internal energy change at each RI collision.
 The analytical results, ``Theory" (solid line) generated from  Eqs. (\ref{eq:workstep_analytical})-(\ref{eq:energystep_analytical}) perfectly agree with the numerical simulations (triangles) obtained from the definitions  (\ref{eq:workstep})-(\ref{eq:Esstep}).
The dashed purple line on zero directs the eye. 
Parameters are $J_{xx}=2,J_{yy}=1,J_{zz}=4,\omega_{A}=2,\omega_{S}=1,\tau=0.01,\beta=1$. The initial state of the  system  is characterized by $c_S^{(0)}=0.459-0.152i$ and $p_S^{(0)}=0.627$.}
    \label{fig:heat-work-energy-numerics}
\end{figure}

In Fig. \ref{fig:heat-work-energy-numerics}, we present the theoretical predictions for the change of energy between work, heat, and the system, obtained from Eqs. (\ref{eq:workstep_analytical}), (\ref{eq:heatstep_analytical}) and (\ref{eq:energystep_analytical}), respectively, and we compare them to direct simulations, showing perfect agreement. 
As the system approaches its steady state, $\Delta E_S^{(n)}$ converges to 0, while $Q^{(n)}$ and $W_{I}^{(n)}$ achieve opposite values, which are the housekeeping work and heat required to sustain the nonequilibrium steady state.

We now provide the formal proof that $\Delta E_{S}^{(\infty)}  = 0$. Let $A$ and $B$ be defined as
\bea
A=\frac{4\omega_{S}(J_{xx}+J_{yy})^{2}}{\theta^{2}}\sin^{2}\left(\frac{\theta\tau}{2}\right), \,\,\,\,\,\,\,
B=\frac{4\omega_{S}(J_{xx}-J_{yy})^{2}}{\phi^{2}}\sin^{2}\left(\frac{\phi\tau}{2}\right);
    \label{eq:AB}    
\eea
Then, it is possible to express the change in the energy of the system as 
\begin{equation}
    \Delta E_{S}^{(n+1)} = (A+B)p_S^{(n)} + [p_{A}(B-A)-B].
    \label{eq: DeltaES_lim_1}
\end{equation}
Using the ansatz for populations, Eq. (\ref{eq:ansatz}), we write
\begin{equation}
    \Delta E_{S}^{(n+1)} = (A+B)\eta^{n}\left(p_S^{(0)}-p_{S}^{(\infty)}\right) + (A+B) p_{S}^{(\infty)} - [B(1-p_{A})+Ap_{A}].
    \label{eq: DeltaES_lim_2}
\end{equation}
Noting that
\begin{equation}
    (A+B) p_{S}^{(\infty)} = [B(1-p_{A})+Ap_{A}],
    \label{eq: DeltaES_lim_3}
\end{equation}
it is easy to check that Eq. (\ref{eq: DeltaES_lim_3}) yields
\begin{equation}
    \Delta E_{S}^{(n+1)} = (A+B)\eta^{n}\left(p_S^{(0)}-p_{S}^{(\infty)}\right).
    \label{eq: DeltaES_lim_4}
\end{equation}
In the limit $n\rightarrow+\infty$, assuming $|\eta|<1$, which holds except for a countable set of points,
 $\Delta E_{S}^{(n)}\rightarrow0^{+}$.
Since $\Delta E_{\text{tot}}^{(n)} = 0$ for every $n$, we have that
\begin{equation}
    \Delta E_{S}^{(n\to \infty)}\rightarrow0^{+}  \Rightarrow Q^{(\infty)}=-W_I^{(\infty)}.
    \label{eq:work_heat}
\end{equation}

\subsection{Examples}
\label{sec:examples}

We simulate the number of RI steps and work input required for building the steady state using different metrics for measuring the proximity to the steady state. Simulations consider both diagonal and nondiagonal initial states of the system, as well as the two interaction Hamiltonians discussed in this work. Additionally, depending on the system's effective temperature, the process may correspond to either a cooling or a heating task.

The calculation of $n^*$ was discussed in Secs. \ref{sec:T} and \ref{sec:F}. Since a general analytic expression for $n^{*}$ is unavailable, we proceed to compute the work required to reach the steady state numerically with the following steps: (i) We compute the state of the system as it evolves in time and its steady-state value either from the closed-form expressions or numerically. (ii) We calculate the number of iterations $n^{*}$, necessary to bring the system $\epsilon$ close to steady state. This number is found by solving numerically Eqs. (\ref{eq:traceeps}) or (\ref{eq:fidelityeps}), using trace distance or fidelity as metrics, respectively.  (iii) We compute the work invested in each RI step using Eq. (\ref{eq:workstep}), add its contributions up to $n^*$, to obtain the total work, and include a negative sign according to our sign convention to indicate the amount of work invested. 

\begin{figure}[htbp]
    \centering
    \includegraphics[width=0.8\linewidth]{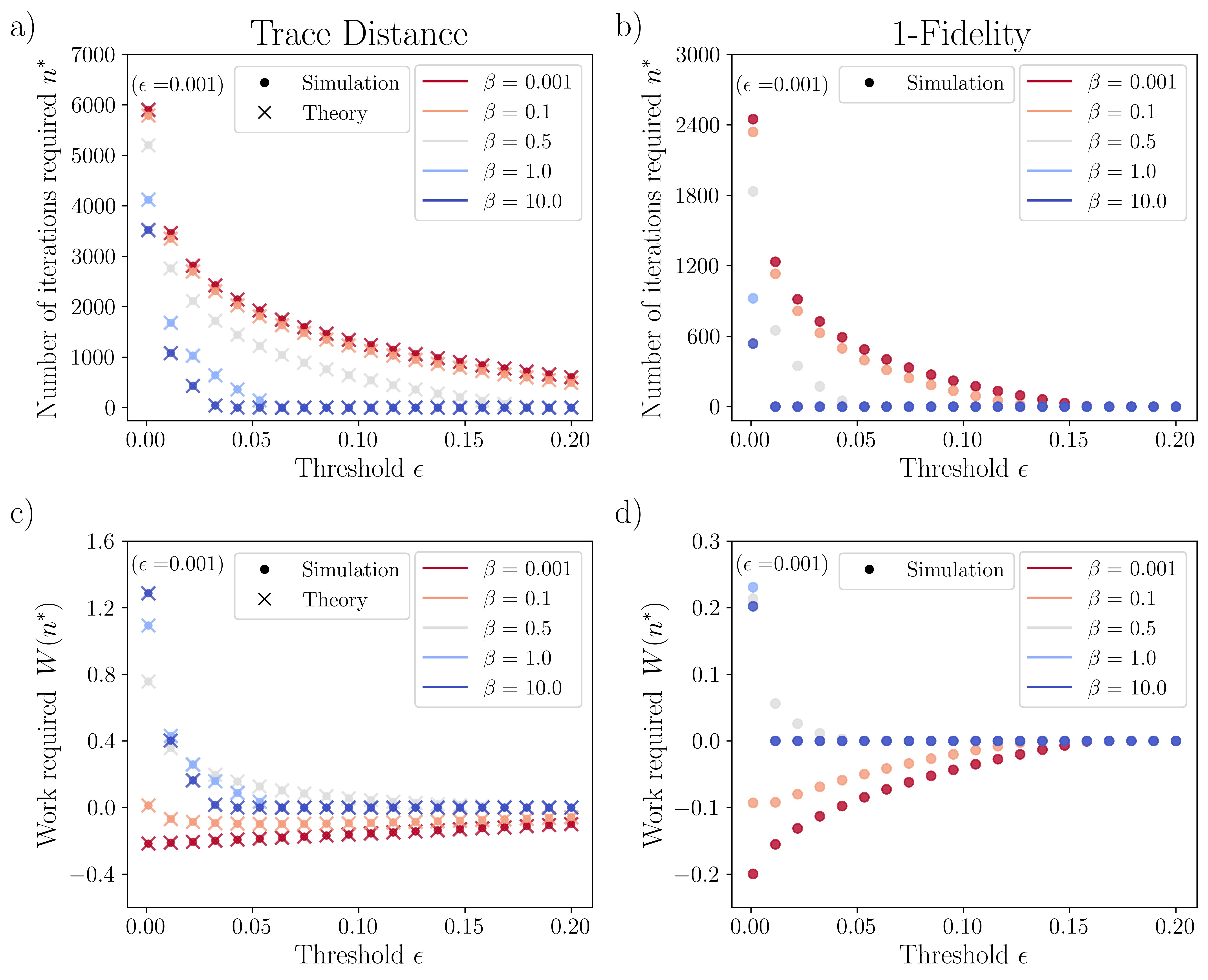}
\caption{Cost-accuracy tradeoff in the relaxation dynamics of the $J_{xx}-J_{yy}$ model towards a nonequilibrium steady state using a {\it diagonal} initial state for the qubit with a ground state population $p_S^{(0)}=0.866$, thus, an effective temperature of $\beta_S=1.866$. The inverse temperatures of the ancilla are $\beta \in \{0.001, 0.1, 0.5, 1.0, 10\}$. 
(a)-(b) Minimum number of RI steps $n^*$ necessary for the system to approach the target state  (\ref{eq:psss}) within a threshold $\epsilon$ based on the trace distance (a) and infidelity (b). 
(c)-(d)  Work cost after $n^*$ RI steps.
Parameters are $J_{xx}=2$, $J_{yy}=1$, $J_{zz}=0$, $\omega_{A}=2$, $\omega_{S}=1$, $\tau=0.01$.} 
\label{fig:ED1}
\end{figure}

\subsubsection{ $J_{xx}-J_{yy}$ model with diagonal $\rho_S^{(0)}$}

In Fig. \ref{fig:ED1}, we assess the cost of reaching steady state assuming that the initial condition of the system is diagonal: the effective initial inverse temperature of the system is $\beta_S = 1.866$, corresponding to the ground state population $p_S^{(0)}=0.866$.
Varying the temperature of the ancilla, either a cooling or heating process of the system is observed. We use both the trace distance (panel a) and the infidelity, $1-F$ (panel b), to compute the required number of iterations, $n^{*}$.
The asymptotic state of the system is given by Eq. (\ref{eq:ptauss}); with our parameters,  $p_S^{(\infty)}= (8p_A+1)/10$. Thus,
if the ancilla is set at low temperature, namely $\beta\gg1$, then $p_A \to 1$, $p_S^{(\infty)}\to 0.9$, and few or no iterations are required to complete the relaxation task as we increase $\epsilon$. 
In contrast, at high temperature of the ancilla, $p_A\to 0.5$ and $p_S^{(\infty)}\to 0.5$. Then, the RI scheme heats up the system, demonstrating a negative energetic cost because of the choice of initial conditions.

Some general observations include:
(i) The two metrics yield different values for $n^*$, with the fidelity measure providing a smaller estimate for $n^*$ compared to the trace distance. This difference translates into less work being consumed in the RI process when estimating accuracy with the fidelity measure.
(ii)  $n^*$ grows monotonically when $\epsilon$ is reduced, which is expected.
(iii) Depending on the temperature of the ancilla relative to both initial and asymptotic state of the system, the task of approaching the steady state concerns either cooling or heating processes with work invested or released in the collision process.
(iv) Having observed that the interaction term $\propto \hat{\sigma}_z^S\otimes\hat{\sigma}_z^A$ only affects coherences, leaving population unchanged, the results reported in Fig. \ref{fig:ED1} are the same for $J_{zz}\neq0$. 

\begin{figure}[htbp]
    \centering
    \includegraphics[width=0.7\linewidth]{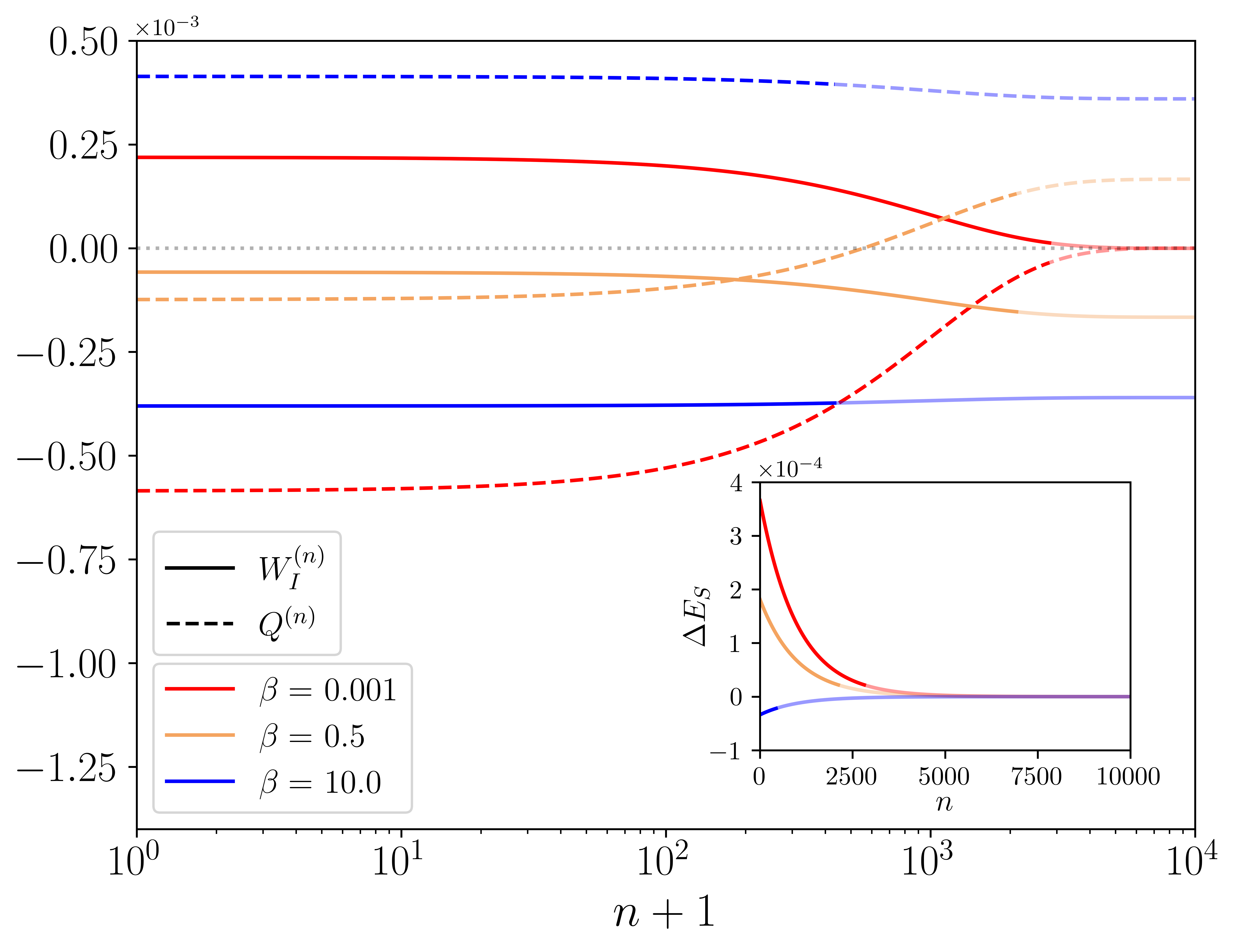}
    \caption{Heat $Q^{(n)}$ and work $W_I^{(n)}$ at each RI step. 
    The interaction Hamiltonian considered is of $J_{xx}-J_{yy}$ type, and the initial state of the system is diagonal at the inverse effective temperature $\beta_S^{(0)} = 1.866$, with $p_S^{(0)} = 0.866$. 
    The full color represent the dynamics until $n^*$ using $\epsilon=0.022$; the faded color stands for the rest of the dynamics, from $n^*$ to the end of simulation. 
 (a) The system's internal energy variation for each step of the RI dynamics. 
 Parameters are $J_{xx} = 2$, $J_{yy} = 1$, $J_{zz} = 0$, $\omega_{S} = 1$, $\omega_{A} = 2$, $\tau = 0.01$, $\beta\in{0.001,0.5,10}$.} 
    \label{fig:ED1_thermo}
\end{figure}

To better understand trends in Fig.\ref{fig:ED1}, we present in Fig. \ref{fig:ED1_thermo}, the changes in work, heat, and the internal energy of the system as a function of the iteration index, $n$.
These thermodynamic quantities have been evaluated from their definitions, Eq. (\ref{eq:workstep}), (\ref{eq:Qstep}), (\ref{eq:Esstep}), as the RI dynamics is iterated. The calculation was repeated for different temperatures, 
$\beta\in\{0.001,0.5,10\}$. 
The main plot shows that $Q^{(\infty)} = -W_{I}^{(\infty)}$ once $n\to\infty$, while the inset confirms that $\Delta E_S^{(n)}\to 0^{+}$, as expected. We note also that for high environmental temperature ($\beta = 0.001$), heat and work in steady state tend to 0, recovering an equilibrium steady state at the ancilla temperature (thermalization). 

\begin{figure}[htbp]
\includegraphics[width=1\linewidth]{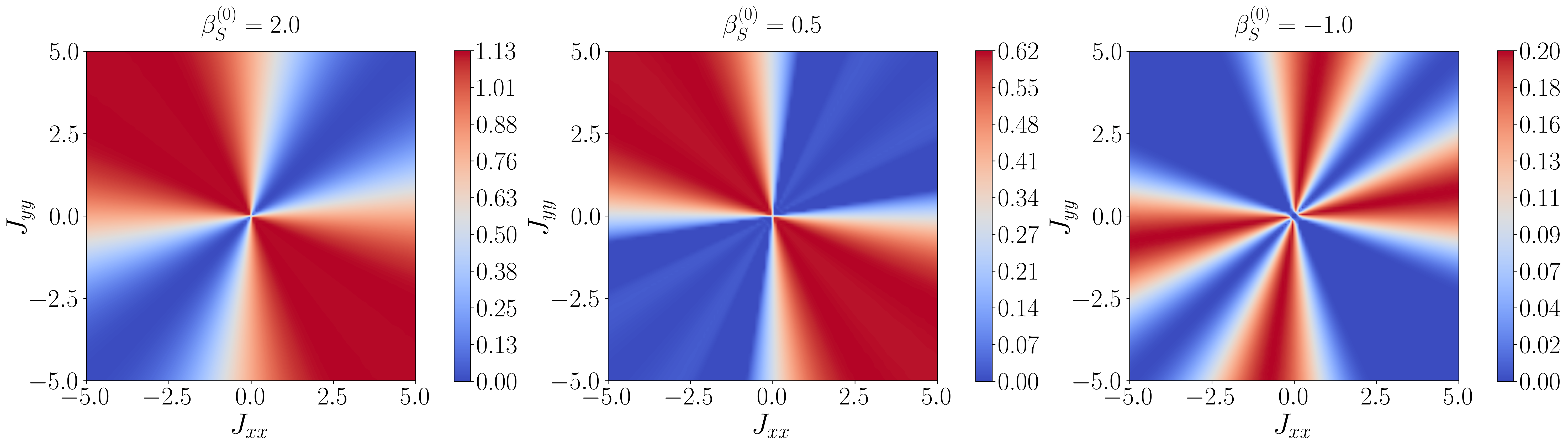} 
\caption{Work cost $W(n^*)$ as a function of $J_{xx}$ and $J_{yy}$ for three diagonal initial conditions for the system, indicated by the effective temperatures in panels (a) -(c).
We use the trace distance metric to calculate $n^*$ through Eq. (\ref{eq:nstar}) with a threshold $\epsilon=0.05$. Parameters are $\omega_A = \omega_S = 1$, $\tau = 0.01$, and $\beta=1$ for the ancilla.}
\label{fig:work_vs_JxxJyy_panel}
\end{figure}

\begin{figure}[htbp]
    \centering
    \includegraphics[width=0.8\linewidth]{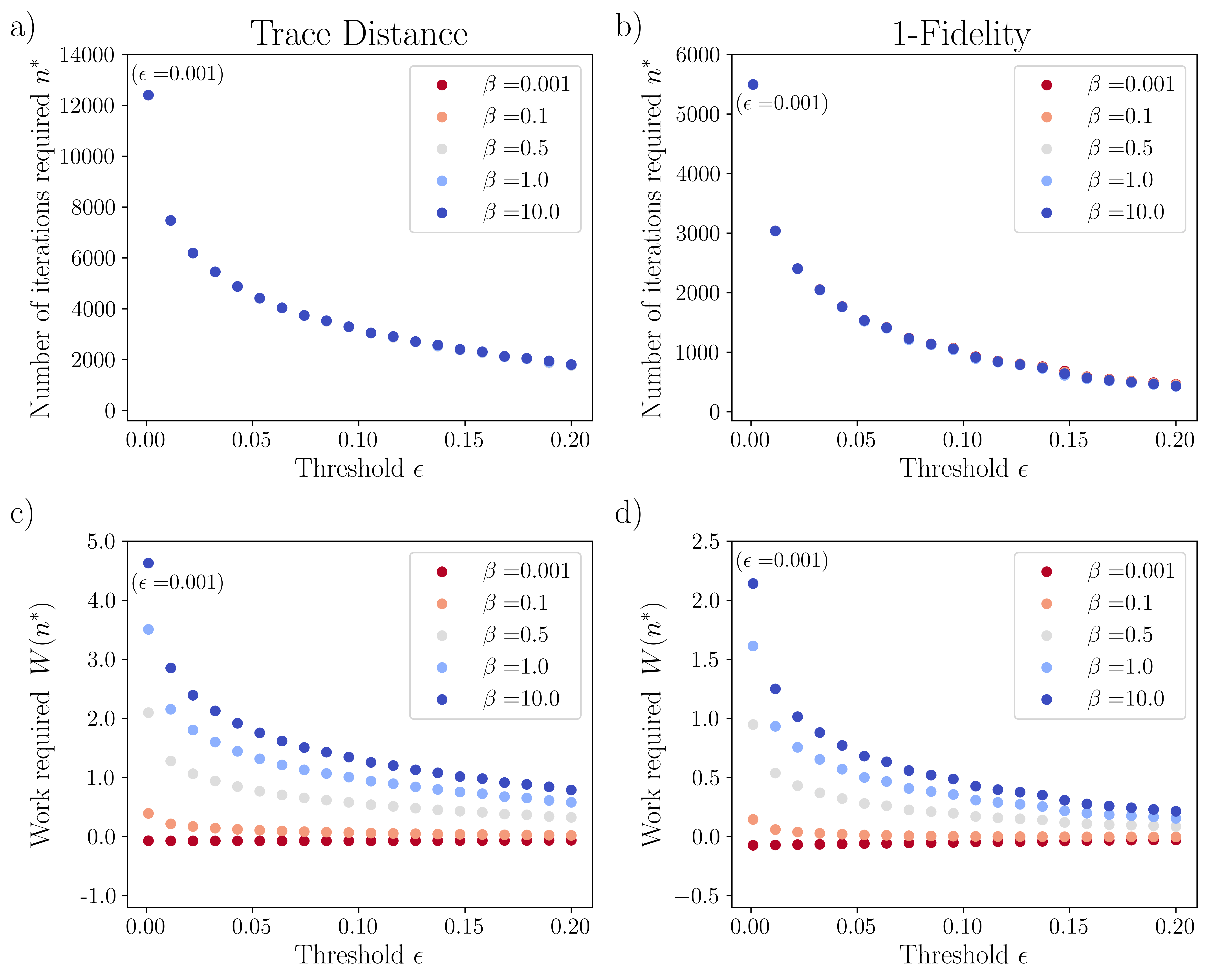} 
\caption{Cost-accuracy tradeoff in an RI process of the $J_{xx}-J_{yy}$ model with a {\it nondiagonal} initial state for the system.
The inverse temperature of the ancilla is $\beta \in \{0.001, 0.1, 0.5, 1, 10\}$.
(a)-(b) Minimum number of RI steps $n^*$ required to approach the target state of the system (\ref{eq:psss}) within a threshold $s\epsilon$ based on the trace distance (a) and infidelity (b).
(c)-(d) Associated work cost after $n^*$ collisions.
Parameters are $J_{xx}=2$, $J_{yy}=1$, $J_{zz}=0$, $\omega_{A}=2$, $\omega_{S}=1$, $\tau=0.01$.}
   \label{fig:END1}
\end{figure}

 
In Figure \ref{fig:work_vs_JxxJyy_panel} we present the work cost for generating the steady state when varying $J_{xx}$ and $J_{yy}$. We assume that the initial state of the system is diagonal, and we test three cases with different effective initial temperatures for the reduced system. These temperatures are defined on the basis of the excited-to-ground-state population ratio.
For every choice of coupling parameters (i) we compute the steady state of the system, which generally differs from the ancilla's state, (ii) we compute the trace distance and find $n^*$ from Eq. (\ref{eq:nstar}), and
(iii) we calculate the work in each RI step, then the accumulated work up to $n^*$ using Eqs. (\ref{eq:workstep})-(\ref{eq:totalwork}).
In all cases, the work cost is zero when $J_{xx}=J_{yy}$, a limit corresponding to ``energy conservation" \cite{RevRI}. In this case, the process corresponds to thermalization with only heat exchanged between the ancillas and the qubit system.
We further find that, if the qubit is prepared in a Gibbs-like thermal state as in Fig. \ref{fig:work_vs_JxxJyy_panel}(a)-(b), more work is invested when $J_{xx}$ and $J_{yy}$ have opposite signs, since under these parameters,  population inversion takes place with $p_S^{(\infty)}<1/2$. In Fig. \ref{fig:work_vs_JxxJyy_panel}(c) we also test the situation where the system is prepared with population inversion described by a negative temperature. 
In this case, we pay no energy on the off-diagonal (and little energy in the vicinity), since the system is already in (or close to) its steady state.


\subsubsection{ $J_{xx}-J_{yy}$ model with nondiagonal $\rho_S^{(0)}$}

We present here the energetic analysis for a system initialized in a state with coherences and evolving under the RI protocol up to its nonequilibrium steady state. In particular, we prepare the initial state with $c_S^{(0)}=0.459-0.152i$ and 
$p_S^{(0)}=0.627$.
The relaxation process, along with its associated energetic cost, now involve adjusting the populations to their steady-state values and decaying of coherences to zero. From the results presented in Fig. \ref{fig:END1}, we observe the following:
The number of steps required to converge within $\epsilon$ to steady state is independent of temperature, for both trace distance and infidelity measures. This can be rationalized by noting that in the $J_{xx}-J_{yy}$ model the decay process of {\it coherences}, down to zero, does not depend on the temperature of the ancilla, see Eq. (\ref{eq:psi}). Since this decoherence process requires more steps than population relaxation, it dictates the number $n^*$, making it independent of temperature. The work required to achieve the steady state depends instead on the temperature of the ancilla, as shown by Eq. (\ref{eq:workstep}).

A similar analysis can be performed to evaluate the work cost of reaching steady state in the $J_{xx}-J_{yy}-J_{zz}$ model. In this case, the decay of coherences is temperature-dependent, although the overall trends in the cost analysis remain similar. An illustrative example is provided in Appendix \ref{appendC}.

\section{Summary and open questions}
\label{sec:summary}
Our work contributes to the theoretical understanding of thermalization, or more generally, creating a nonequilibrium steady state within the repeated interaction model. We extended the scope of previous studies by considering a broader class of interactions, allowing for arbitrary coupling parameters $J_{xx}$, $J_{yy}$, as well as $J_{zz}$, at arbitrary interaction strength and duration.

Through analytical derivations, we characterized the behavior of populations under the RI map: We proved that population evolve independently of coherences to their steady state. We showed that populations converge to a fixed-point solution regardless of the initial conditions, and we explicitly determined this solution. For coherences, while we were unable to provide a general proof of their decay, we offered supporting evidence---both analytical and numerical---suggesting that the system converges to a fixed point with zero coherences in our models. Remarkably, for both models investigated in this work, 
the $J_{xx}-J_{yy}$ and the $J_{xx}-J_{yy}-J_{zz}$,  the steady state of the system was diagonal in its energy eigenbasis. However, this steady state depended not only on the temperature of the ancilla as in standard thermal states, but also on the interaction parameters and, notably, the RI timestep---if the latter was not sufficiently small.

Although generically our model Hamiltonian created a non-thermal nonequilibrium steady state that depends on the interaction parameters, we identified from our analytical solution an RI protocol that allowed thermalization of the system with the ancillas. Notably, it involves collisions of long duration and weak (potentially random) interactions.
Moreover, only a few such long and weak collisions sufficed for thermalization. 

In addition to investigating the process of reaching the steady state from generic initial conditions, we estimated the associated resources: runtime and energy cost, the latter evaluated by calculating the work required for the process. 
For diagonal initial states of the system, we used the trace distance measure and found a lower bound on the number of RI collisions required to get $\epsilon$ close to the steady state. This number scaled logarithmically with the desired accuracy, $\epsilon$, and inversely with the population relaxation rate $|\eta|$.
Furthermore, as expected, we observed a monotonic increase in the work as the convergence threshold to approach the steady state was tightened. The required work was found to depend on both the system's initial state and the interaction parameters.

Although the RI scheme is conceptually straightforward to implement, open questions remain: What type of dynamics can the RI model encompass and what characterizes the resulting steady states? 
Could more advanced RI schemes, such as those incorporating randomization, facilitate {\it thermalization} of more general models to the ancilla's Gibbs thermal state?
Future research will aim to extend the RI framework to multilevel systems and refine energy cost estimations, particularly in the context of quantum algorithms.

\begin{acknowledgements}
We acknowledge fruitful discussions with Matthew Pocrnic and Matthew Hagan. M. F. was supported by an NSERC USRA grant and the University of Toronto Physics Chair's Scholar funding.
D.S. acknowledges the NSERC Discovery Grant and the Canada Research Chairs Program.
A. P. work was supported by the Department of Physics at the University of Toronto and by the research project: ``Quantum Software Consortium: Exploring Distributed Quantum Solutions for Canada" (QSC). QSC is financed under the National Sciences and Engineering Research Council of Canada (NSERC) Alliance Consortia Quantum Grants \#ALLRP587590-23.
\end{acknowledgements}

\appendix
\section{Proof of $-1<\eta\leq 1$} 
\label{appendA}

In this Appendix, we provide an analytical proof for the bounds $-1<\eta\leq 1$. 
This result is essential in order to demonstrate that the ground-state population of the reduced system  converges geometrically 
to a fixed steady-state solution, as given by Eq. (\ref{eq:psss}). 
We begin our analysis with Eq. (\ref{eq:eta2}), rewritten as
\bea
\eta=1-M_{\theta}-M_{\phi},
\eea
where
\bea
M_{\theta}&=&        \frac{4(J_{xx}+J_{yy})^2}{\theta^{2}} \sin^{2}\left(\frac{\theta\tau}{2}\right), 
\nonumber\\
 M_{\phi}&=&        \frac{4(J_{xx}-J_{yy})^2}{\phi^{2}}\sin^{2}\left(\frac{\phi\tau}{2}\right).
\eea
Since $0\leq M_{\theta}\leq1$ and $0\leq M_{\phi}<1$, it is trivial to note that $\eta>-1$ since this is equivalent to $M_{\theta}+M_{\phi}<2$;
recall the definition of the energy parameters $\theta$ and $\phi$ given in Eq. (\ref{eq:thetaphi}). We also note that $\eta\leq 1$, since this corresponds to $M_{\theta}+M_{\phi}\geq0$, which is always true. In particular, it can be shown that the rate is smaller than one, $\eta<1$, except at special points discussed next. 

Consider the $\eta=1$ case. This means that $M_{\theta} = -M_{\phi}$.
Since the left-hand side and right-hand side have opposite signs, the equality can only be satisfied if both terms are simultaneously zero,
\begin{eqnarray}
    \tau\theta =  2\pi m \Rightarrow \tau = \frac{ 2\pi m}{\theta} \:(m \in \mathbb{Z}) \\
    \tau\phi = 2\pi k \Rightarrow \tau = \frac{2\pi k}{\phi} \: (k \in \mathbb{Z}).
\end{eqnarray}
These conditions (i) allow $\tau$ to take particular solutions and (ii)  enforce $\theta = q\phi, \:\: q \in \mathbb{Q}$, which is satisfied for a combination of $J_{xx},J_{yy},\omega_{S},\omega_{A}$ such that
\begin{equation}
    \frac{\theta^{2}}{\phi^{2}} = \frac{4(J_{xx}+J_{yy})^{2}+(\omega_{A}-\omega_{S})^{2}}{4(J_{xx}-J_{yy})^{2}+(\omega_{A}+\omega_{S})^{2}} = q^{2}, \;\; q\in\mathbb{Q}.
    \label{eq:timeA}
\end{equation}
For example, if $J=J_{xx}=J_{yy}$ and $\omega=\omega_A=\omega_S$, then $\theta= 4 J$ and $\phi=2 \omega$.
The above conditions thus translate to  
$2 \omega\tau= 2\pi k$ and $4 J \tau = 2\pi m$.
Selecting, for example, $J=\omega=1$, we find that $\eta=1$ when the time step is set to
$\tau=k\pi$ with $m=k/2$.

These special time steps that satisfy (\ref{eq:timeA}) could be regarded as specific Rabi oscillation periods in which the interaction process is not impactful because the state of the system returns to its original value within the collision time. For these specific solutions, the system maintains its initial condition indefinitely. 

However, the RI scheme is commonly used when $\tau\theta\ll1$ and $\tau \phi\ll1$, in which case the steady-state result becomes independent of $\tau$.  Consequently, for any relevant physical scenario with a short timestep as mentioned above, we get $-1<\eta<1$ and that the qubit populations converge to their fixed point geometrically.


\section{Decay of coherences}
\label{appendB}

As stated in the main text, providing an analytical proof for $|\psi|<1$ is cumbersome. Thus,  this statement has been verified numerically by randomly sampling the parameter space. In particular, we pick $10^{8}$ sets of random values for the parameters that enter  Eq. (\ref{eq:psi}) according to the following probability distributions: $J_{xx}$ and $J_{yy}$ are taken from a uniform distribution $U(-100,100)$, $\omega_{A}$ and $\omega_{S}$ from $U(0,100)$, the phase $\chi$ from $U(-100,100)$, the collision time $\tau$ from $U(0,100)$, and the initial ground state population of the ancilla $p_{A}$ from $U(0,1)$. 
Evaluating Eq. (\ref{eq:psi}) for this collection of examples, we confirmed that $|\psi|<1$ was satisfied.

As for an analytic proof, in the limit of small collision time, $\tau\rightarrow0^{+}$, it is easy to show that $|\psi|\leq 1$ analytically: Taylor-expanding the expression for $\psi$, Eq. (\ref{eq:psi}), to second order in $\tau$ around $\tau = 0$ provides
\begin{equation}
\label{psiTaylor}
\hspace{-1cm}
\psi = e^{i\chi}(1-i\omega_{S}\tau) + \left[\left(J_{xx}^{2}-J_{yy}^{2}\right)e^{-i\chi}-\frac{1}{8}\left(\theta ^2 + \phi ^2 - 2(\omega_{A} ^2 - \omega_{S} ^2)\right) e^{i\chi} \right]\tau ^2 + O(\tau^{3}).
\end{equation}
Calculating $|\psi|^{2} = \psi\psi^{*}$ and neglecting terms in $\tau$ of higher order than two, we get
\bea
    |\psi|^{2} &=& 1 + \left[2(J_{xx} ^2 - J_{yy} ^2)\cos(2\chi)-\frac{1}{4}\left(\theta ^2+\phi ^2 -2(\omega_{A} ^2 - \omega_{S} ^2)\right) + \omega_{S}^2 \right]\tau^{2} + O(\tau ^3) 
    \nonumber\\
    &=& 1 + 2\left[\left(J_{xx} ^2 - J_{yy} ^2\right)\cos(2\chi)-J_{xx}^{2}-J_{yy}^{2}\right]\tau^{2} + O(\tau ^3) 
    \nonumber\\
    &\leq&  1 + 2\left[(J_{xx} ^2 - J_{yy} ^2)-J_{xx}^{2}-J_{yy}^{2}\right]\tau^{2} + O(\tau ^3) 
    \nonumber\\
    &=& 1-4J_{yy}^{2}\tau^{2} + O(\tau ^3).
\label{eq:psimax}   
\eea
This inequality was reached by using  Eq. (\ref{eq:thetaphi})
and substituting in the third line $\cos(2\chi) \to 1$. The latter replacement corresponds to maximizing the expression with respect to $\chi$. 
Together,
\begin{equation}
    \label{psilimit}
    |\psi|^{2} \approx 1-4J_{yy}^{2}\tau^{2} \leq1 \:\:\:\:\:\:as\:\: \tau \approx 0.
\end{equation}
In the limit of short collision time,  $|\psi|<1$ as long as $J_{yy}\neq0$.
If $J_{yy}=0$, it is possible to look for an analytical proof of $|\psi|<1$ by expanding (\ref{eq:psi})
to the third order in $\tau$, around $\tau=0$. 
However, experimenting in this direction did not prove fruitful. We obtained another term inside Eq. (\ref{psiTaylor}), which was not always negative. 
It remains important to complement numerical simulations and approach this problem analytically to prove that $|\psi|^2<1$ so as to guarantee the decay of coherences to zero inside the RI scheme. In the meantime, in this work we verified this behavior with extensive numerical simulations.

\begin{figure}[h]
    \centering
    \includegraphics[width=0.8\linewidth]{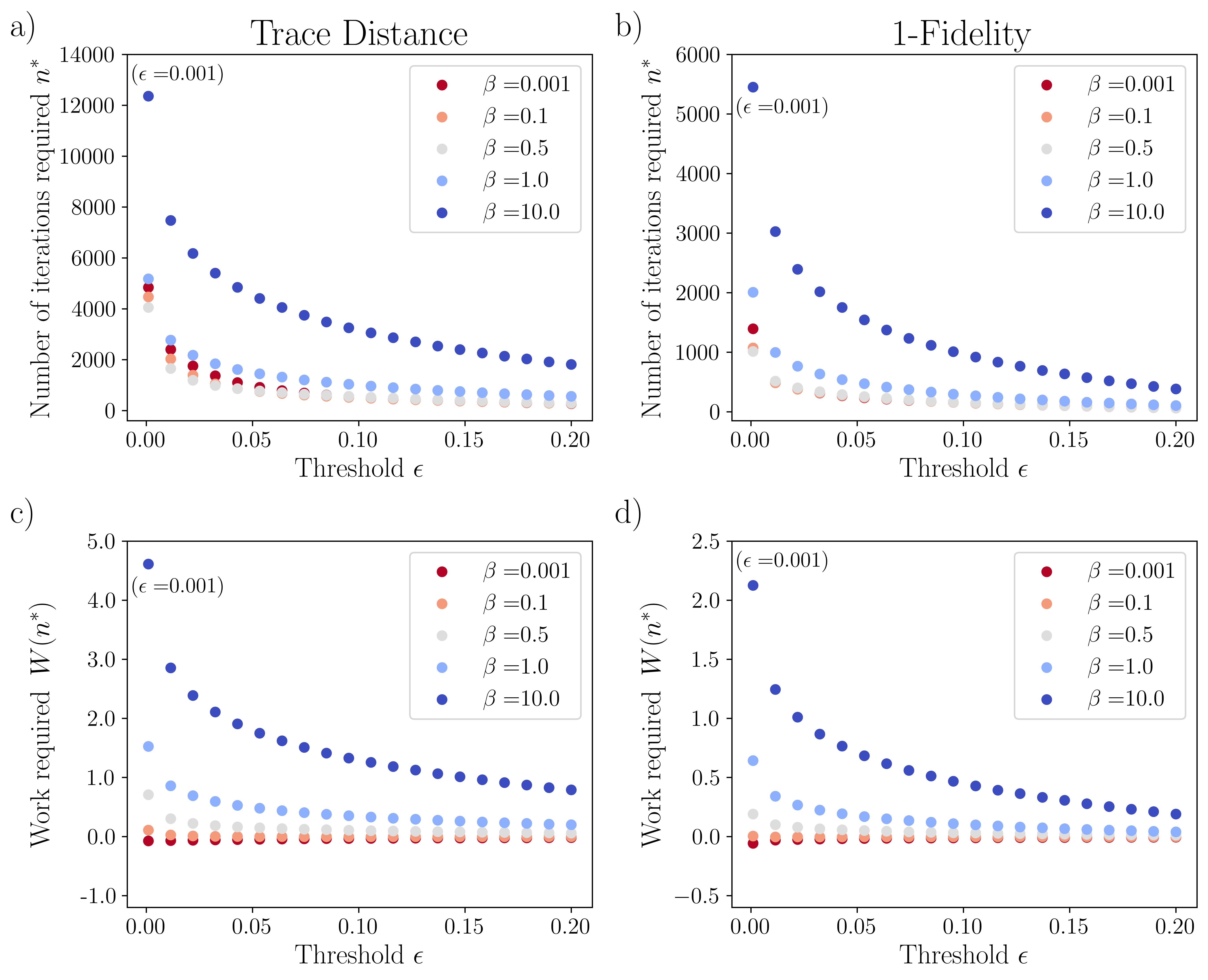} 
    \caption{Cost-accuracy tradeoff in the RI relaxation process of the $J_{xx}-J_{yy}-J_{zz}$ model using a {\it non diagonal random} initial state the same used for Fig. \ref{fig:END1}, prepared with $c_{S}^{(0)}= 0.459-0.152i$ and $p_{S}^{(0)} = 0.627$. The inverse temperatures of the ancilla are $\beta \in \{0.001, 0.1, 0.5, 1.0, 10\}$. (a)-(b) Minimum number of RI steps $n^*$ necessary for the system to approach the target state  (\ref{eq:psss}) within a threshold $\epsilon$ based on the trace distance (a) and infidelity (b). (c)-(d)  Work cost after $n^*$ RI steps. Parameters are $J_{xx}=2$, $J_{yy}=1$, $J_{zz}=4$, $\omega_{A}=2$, $\omega_{S}=1$, $\tau=0.01$.}
    \label{fig:NDS2}
\end{figure}


\section{Energetic cost for the Heisenberg Hamiltonian} 
\label{appendC}

We present here  the energetic cost associated with the relaxation dynamics in the RI scheme in the $J_{xx}-J_{yy}-J_{zz}$ model,
complementing the main text where we studied the   $J_{xx}-J_{yy}$ model for diagonal and nondiagonal initial conditions, Figs. \ref{fig:ED1} and Fig. \ref{fig:END1}, respectively.

We adopt here the same nondiagonal state used in Fig. \ref{fig:END1}, but now evolving it under the $J_{xx}-J_{yy}-J_{zz}$ Hamiltonian.
Comparing Fig. \ref{fig:NDS2} to Fig. \ref{fig:END1}, we find that now $n^*$ depends on temperature. This is to be expected from our previous analysis, Fig. \ref{fig:beta}, where we showed that the decoherence process depended on temperature when $J_{zz}\neq 0$.



\begin{thebibliography}{4}
\bibitem{commentT}
We refer to the process of the system reaching a diagonal state with
populations equal to that of the ancilla's, $p_S=p_A$, as ``thermalization". In this steady state solution, the system achieves the same diagonal state as the bath's, and, as we show in Eq. (\ref{eq:workstep_analytical}), there is no housekeeping work involved in maintaining this Gibbs state, which is thus an equilibrium solution.
\end{thebibliography}
\end{document}